\documentclass[%
reprint,
amsmath,amssymb,
aps,
pre,
]{revtex4-2}

\usepackage{graphicx}
\usepackage{dcolumn}
\usepackage{bm}
\usepackage{hyperref}
\usepackage{comment}
\usepackage{mathtools}

\DeclareMathOperator{\Softmax}{Softmax}

\DeclareMathOperator{\bxi}{\boldsymbol{\xi}}
\DeclareMathOperator{\bx}{\mathbf{x}}
\DeclareMathOperator{\alb}{\alpha_{c}^{lb}}
\DeclareMathOperator{\albr}{\alpha_{c,\text{rams}}^{lb}}

\DeclareMathOperator{\ctc}{\cos\theta_c}
\DeclareMathOperator{\tc}{\theta_c}

\DeclareMathOperator{\st}{st}

\DeclareMathOperator{\R}{\mathbb{R}}

\newcommand{\ve}{{\varepsilon}}
\newcommand{\cS}{{\mathcal{S}}}

\begin{document}
	
\title{The Exponential Capacity of Dense Associative Memories }
\author{Carlo Lucibello}
\email{carlo.lucibello@unibocconi.it}
\author{Marc Mézard}%
\email{marc.mezard@unibocconi.it}
\affiliation{%
Department of Computing Sciences, Bocconi University, Milano, Italy\\
Bocconi Institute for Data Science and Analytics (BIDSA), Milano, Italy
}%
	
\begin{abstract}
Recent generalizations of the Hopfield model of associative memories are able to store a number $P$ of random patterns that grows exponentially with the number $N$ of neurons, $P=\exp(\alpha N)$. Besides the huge storage capacity, another interesting feature of these networks is their connection to the attention mechanism which is part of the Transformer architectures widely applied in deep learning.
In this work, we study a generic family of pattern ensembles using a statistical mechanics analysis which gives exact asymptotic thresholds for the retrieval of a typical pattern, $\alpha_1$, and lower bounds for the maximum of the load $\alpha$ for which all patterns can be retrieved, $\alpha_c$, as well as sizes of attraction basins.
We  discuss in detail the cases of Gaussian and spherical patterns, and show that they display rich and qualitatively different phase diagrams. 
\end{abstract}

\maketitle

\section{Introduction}
\label{sec:intro}
About forty years ago, John Hopfield introduced a simple model of memory \cite{Hopfield1982} based on $N$ spins interacting by pairs like in a spin glass, but with specifically tailored interactions so that the ground states of the spin-glass are strongly correlated with a set of patterns that one wants to memorize. This allows building an associative memory that retrieves the full information from some partial information. 
Using statistical physics methods, it was then shown that this model can store up to  $\alpha_c N$ independent random patterns in the large $N$ limit, with $\alpha_c\simeq 0.14$ if one uses the simple Hebb rule for defining the interactions \cite{Amit1985aPRA,Amit1985bPRL}.
Going beyond pairwise interactions, i.e. introducing $p$-spin interactions, yields a big increase in the memory capacity which becomes of order $N^{p-1}$ \cite{Gardner1987multiconnected, Krotov2016, agliari2023dense}. Pushing this strategy further, a
family of models with exponential interaction terms leading to exponential capacity has been recently introduced and analyzed in Refs. \cite{Demircigil2017,ramsauer2021hopfield}.
While \cite{Demircigil2017} discusses networks with Ising variables,
Ref. \cite{ramsauer2021hopfield} considers continuous variables and links
the dynamics of the system with the attention mechanism one finds
in the transformer models \cite{vaswani2017attention} now ubiquitous in natural language processing and other domains in machine learning.
We refer to all these generalization as Dense Associative Memories (DAMs).
Here we perform a statistical mechanics analysis 
of the DAM with continuous variables introduced in \cite{ramsauer2021hopfield}. We expect that a similar approach can be used in the case \cite{Demircigil2017} of discrete variables and other DAMs. 
Defining $P=e^{\alpha N}$ as the number of patterns, we want to compute the critical value $\alpha_{c}$ such that, for $\alpha<\alpha_{c}$ retrieval is possible with high probability in the large $N$ limit. 
Notice that the precise definition of $\alpha_{c}$ in presence of an exponentially large number of patterns requires some care, and indeed we will find distinct thresholds depending on whether we request to store all patterns, or we request that a randomly chosen pattern can be retrieved with probability one.

Following \cite{ramsauer2021hopfield}, we consider a set of
$P=e^{\alpha N}$ patterns $\{\bxi^{\mu}\}_{\mu=1}^{P}$, 
$\bxi^{\mu}\in\mathbb{R}^{N}$, independently and identically
distributed according to some distribution $\mathcal{P}(\bxi)$, 
normalized such that $\mathbb{E}\lVert\bxi\rVert^2=N$.
These patterns should be memorized by a network of $N$ neurons, with activities encoded in a $N$-dimensional vector $\bx$. Starting from an initial condition $\bx_{0}\in\mathbb{R}^{N}$,
the recall of a memorized pattern is based on a gradient descent of the energy function 
\begin{align}
	E(\bx) & =-\frac{1}{\lambda}\log\sum_{\mu=1}^{P}e^{\lambda\bx \cdot\bxi^{\mu}}+\frac{1}{2}\lVert\bx\lVert^{2},
	\label{eq:enedef}
\end{align} 
Notice that with respect to Ref. \cite{ramsauer2021hopfield} we exchanged the roles of $\bx$ and $\bxi$ in order to conform to statistical physics' standard notation. 
The gradient descent with learning rate one gives:
\begin{align}
	\bx_{t+1}= & \sum_{\mu}a_t^\mu\,\bxi^{\mu}\; , \qquad a_t^\mu =\frac{e^{\lambda \bx_t\cdot\bxi^{\mu}}}{\sum_{\nu}e^{\lambda\bx_t\cdot \bxi^{\nu}}}\; .
	\label{eq:update}
\end{align}
Interestingly, this update rule is well known in 
deep learning \cite{Bahdanau14attention, vaswani2017attention}. It corresponds to a cross-attention mechanism with a single query $\bx=\mathbf{q}$
and identical key and value matrices: $K=V=[\bxi^1,...,\bxi^P]$. The update rule can be compactly written as  $\bx_{t+1} = V\Softmax(\bx_t^T K)$. The coefficients $a_t^\mu$ are called attention scores.

A pattern $\bxi^\mu$ will be said to be retrieved if,
starting from an initial configuration $\bx_{0}$ close enough to the pattern, the gradient iterations converge close to the pattern: $\lim_{t\to\infty}\frac{1}{N} \lVert \bx_t - \bxi^\mu\rVert^2 < \epsilon$ for some constant $\epsilon>0$ that we assume to be vanishing in the limit of large $N$. In the standard Hopfield model with $P=O(N)$ patterns, the retrieved configuration always contains a small fraction of errors compared to the original pattern, and one has to impose $P=O(N/\log(N))$ for perfect retrieval \cite{Bovier99}. For our DAM, we discuss (asymptotic) perfect retrieval.

At fixed interaction strength $\lambda$, we define the single pattern retrieval threshold $\alpha_{1}(\lambda)$ as the largest value of $\alpha$ for which the probability of retrieving a randomly chosen pattern goes to one in the large $N$ limit. 
This is a sharp phase transition: as we shall see, for $\alpha>\alpha_1(\lambda)$ this probability goes to zero. The other problem we consider is the full retrieval problem, that is determining the maximum number of patterns that are jointly stored. We define the capacity threshold $\alpha_{c}(\lambda)$ as the largest $\alpha$ such that all $P=e^{\alpha N}$ patterns are retrieved. Clearly we have $\alpha_c(\lambda) \leq \alpha_1(\lambda)$.

In this paper, we derive exact and simple expressions for $\alpha_1(\lambda)$, provide bounds on $\alpha_c(\lambda)$, and study the size of attraction basins, i.e. the maximal distance between $\bx_0$ and the pattern such that retrieval is possible. We shall first explain the general formalism and then apply it to various distributions of patterns.

\section{Retrieval of a typical pattern}
We study the capacity in the thermodynamic limit $N\to\infty$.
Let us choose one of the patterns
independently from the sample realization, say
$\boldsymbol{\xi}^{1}$,
and explore the energy landscape $E(\bx)$ in its
neighborhood, 
when the overlap   $\bx\cdot\bxi^{1}$, as well as the square norm $\Vert\bx\Vert^2$ are of $O(N)$. The overlaps of $\bx$  with each of the $e^{\alpha N}-1$ other
patterns are typically of order $\sqrt{N}$; however,
as their number is exponentially large, some of these overlaps will
also be of order $N$. 
It is convenient to rewrite the energy as:
\begin{equation}
	E(\bx)=-\frac{1}{\lambda}
	\log\left(e^{\lambda \bx \cdot\bxi^{1}}+e^{\lambda N  \Phi(\mathbf{x})}\right) +\frac{1}{2}\lVert\bx\lVert^{2},
	\label{eq:ene_decomp}
\end{equation}
where 
\begin{equation}
	\Phi(\bx)= \frac{1}{\lambda N}\log \left(
	\sum_{\mu=2}^{P}e^{\lambda\bx \cdot\bxi^{\mu}}\right)\ .
        \label{eq:def-Phi}
\end{equation}
The $\log$ term in the energy is the sum of two terms which are both exponential in $N$. We identify the first term $e^{\lambda\bx \cdot\bxi^{1}} $ as  signal and $e^{N\Phi(\bx)}$ as  noise. In fact, if $\Phi$ can be neglected,  the energy becomes a quadratic function with minimum in $\bxi^1$.

We now study the noise term. For given $\bxi^{1}$ and $\bx$ independent from the other patterns,  $\Phi$
is a random variable whose distribution is induced by the random
patterns $\bxi^{\mu}$ with $\mu=2,...,P$.
Notice that $\Phi$ is minus the free energy density of a
Random Energy Model (REM) \cite{Derrida1981REM} with $P-1$ ``energies"
$\ve^\mu=\bx \cdot\bxi^{\mu} / N$ and ``inverse temperature" $\lambda$. Let us stress that these energies and temperature are auxiliary quantities defined within the REM, they have nothing to do with the energy function or temperature of the associative memory under study. 

As recalled in Appendix \ref{app:noisefunction}, the REM can be solved both with direct probabilistic methods and with the replica method. For the sake of the analysis, we shall focus hereafter on settings where
the distribution of $\ve^\mu$ depends on $\bx$ only through its rescaled norm $\rho=\lVert\bx\rVert /\sqrt{N}$. This is the case for
pattern distributions that are rotationally invariant.   
In the thermodynamic limit where  $N,P\to \infty$ with fixed exponential rate $\alpha$ and considering a sequence of $\bx$ chosen independently from $\{\bxi^\mu\}_{\mu=2}^P$ and at fixed   $\rho=\Vert\bx\Vert/\sqrt{N}$, the quantity $\Phi(\bx)$ converges almost surely to a value
$\phi_{\alpha,\rho}(\lambda)$. This asymptotic free energy can be expressed in terms of the cumulant generating function of the $\ve^\mu$. In the case of Gaussian patterns, it reads:
\begin{align}
\phi_{\alpha,\rho}(\lambda) = \begin{cases}
\frac{\alpha}{\lambda}+\frac{1}{2}\lambda \rho^2 & \lambda<\lambda_*(\alpha,\rho)\\
\rho\sqrt{2\alpha} & \lambda\geq\lambda_*(\alpha,\rho),
\end{cases}
\label{gaussian_Phi}
\end{align}
where $\lambda_*(\alpha,\rho)=\sqrt{2\alpha}/\rho$. The generic expression is given in \eqref{eq:phi_rem-final}.
 In general, $\phi_{\alpha,\rho}(\lambda)$ is  a decreasing function of $\lambda$ and an increasing function of $\alpha$. As exemplified in the Gaussian case \eqref{gaussian_Phi}, it takes two different forms separated by a glass phase transition line in the plane $(\alpha,\lambda)$. For $\lambda <\lambda_*(\alpha,\rho)$, exponentially many energies $\ve^\mu$ contribute to $\Phi(x)$; on the other hand, for $\lambda \geq\lambda_*(\alpha,\rho)$, the REM is dominated by the largest $\ve^\mu$: this is called condensation. 

 We now use this REM analysis in order to study the energy landscape of (\ref{eq:ene_decomp}).
%
For large $N$ we have:
\begin{equation}
	E(\bx) \approx -\max(\bx \cdot\bxi^1,\,N\phi_{\alpha,\rho}(\lambda))+\frac{1}{2}\lVert\bx\lVert^{2}.
	\label{Eres}
\end{equation}
Whenever $\bx \cdot \bxi^1$ dominates the $\max$, the energy $E(\bx)$ is a quadratic well with a minimum in $\bx=\bxi^1$, and the pattern $\bxi^1$ is retrieved in one step of gradient descent. This occurs at small $\alpha$. When the storage $\alpha$ increases, $\phi_{\alpha,\rho}(\lambda))$ increases and the basin of attraction of $\bxi^1$ shrinks. Above a critical value of $\alpha$, the quadratic well disappears, because $E(\bx=\bxi^1)$ is dominated by the noise term. 
We thus have a simple criterion to identify the critical value $\alpha_1$ such that for $\alpha < \alpha_1$ retrieval of $\bxi^1$ is possible:
\begin{equation}
	\alpha_1(\lambda) = \sup \left\{\,\alpha\, :\,  \phi_{\alpha,1}(\lambda)<1\,\right\}.
		\label{eq:alpha1}
\end{equation}
Notice that we set $\rho=1$ since $\mathbb{E}\lVert\bxi^1\rVert^2=N$. 
What happens at $\alpha>\alpha_1$ when the pattern is not retrieved? Ramsauer et al. \cite{ramsauer2021hopfield} show that in this case the attention scores become approximately flat and the dynamics converges to the barycenter of the patterns. In fact, it can be shown that in our high-dimensional regime 
$\nabla E(\bx)$ develops an extensive radial component leading the dynamics towards the origin.

Remarkably, we could derive the exact threshold $\alpha_1$ thanks to a simple inspection of the energy function,
bypassing calculations involving the dynamical rule or even quenched free energy computations à la Ref. \cite{Amit1987AP}. This is a consequence of the exponentially many terms appearing in the expression of the energy, thanks to which the signal-vs-noise balance becomes an all-or-nothing one. 
On the other hand, let us mention a subtlety of the derivation: in the REM analysis of $\Phi(\bx)$, we assumed the choice of $\bx$ to be independent from $\{\bxi^\mu\}_{\mu=2}^P$. Therefore this analysis does not prove that Eq. \eqref{Eres}  holds for {\it all } $\bx$ in a neighborhood of $\bxi^1$.  However, it does hold for $\bx=\bxi^ 1$, and by continuity we can deduce that,
for $\alpha <\alpha_1(\lambda)$, there exists a 
ball centered at $\bxi^1$ with a non-vanishing (and extensive) radius, such that the dynamics converges 
 to the pattern for all $\bx$ inside this ball. 

\section{Retrieval of all patterns}
\label{sec:allpatterns}

\begin{figure}[t]
	\includegraphics[width=0.45\textwidth, trim={5 0 0 25}, clip]{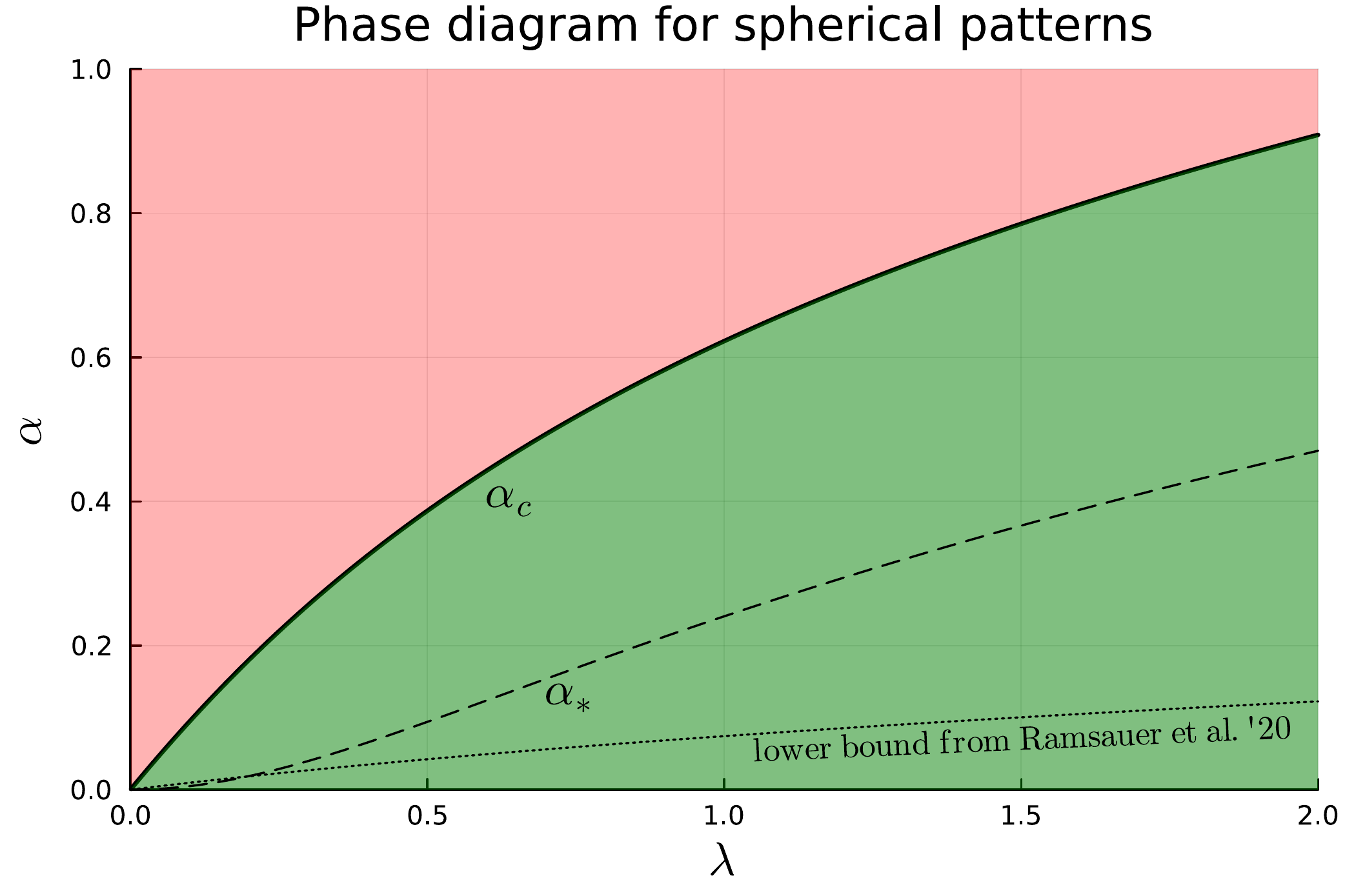}
	\includegraphics[width=0.45\textwidth, trim={5 0 0 25}, clip]{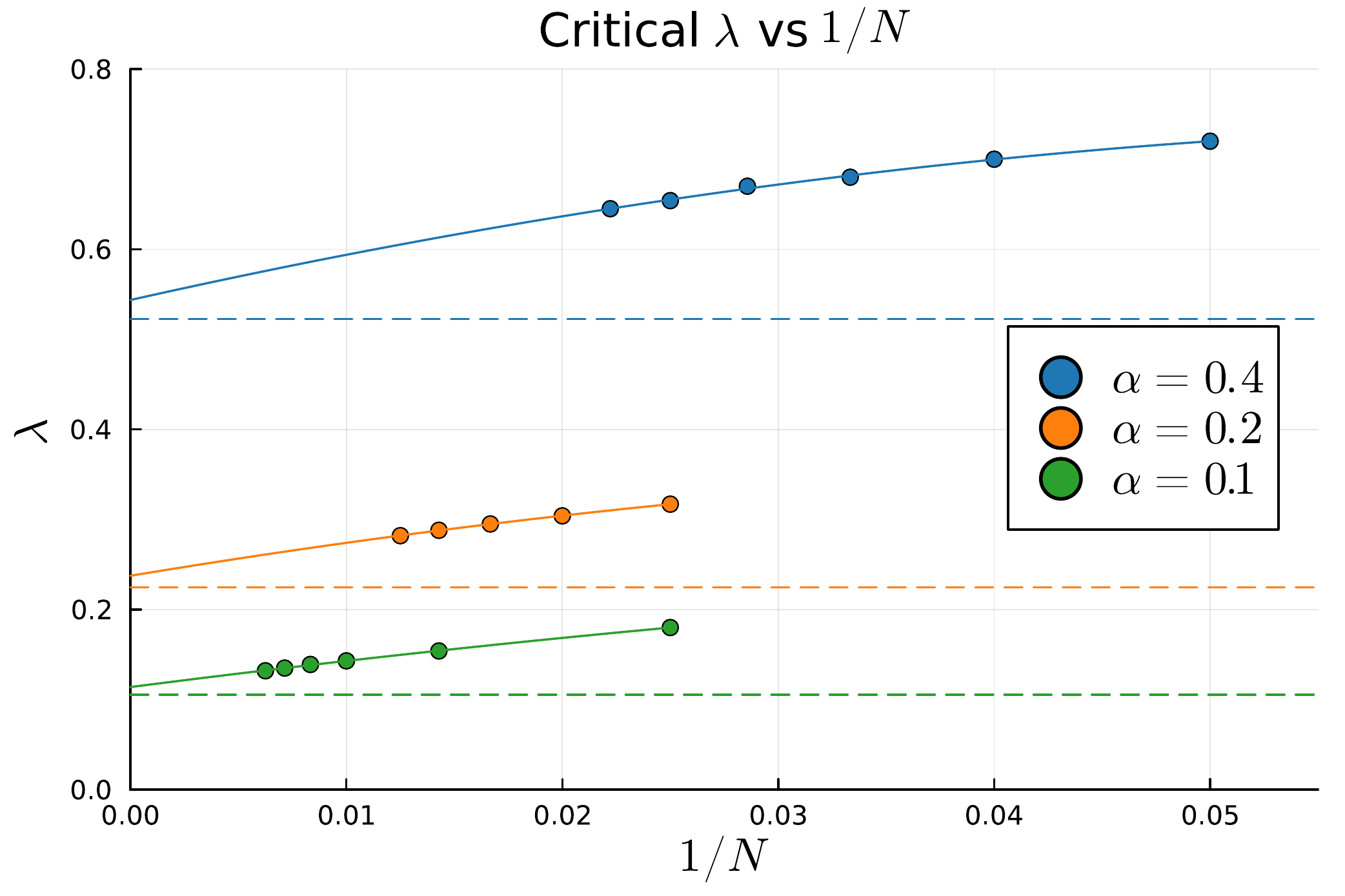}
	\caption{\textbf{Top}: Phase diagram for  patterns uniformly distributed on the hypersphere. All the patterns are retrieved in the green region $\alpha<\alpha_c=\alb=\alpha_1$ (the two thresholds coincide in the spherical setting). The dashed line $\alpha_*(\lambda)$ is where the REM condensation occurs, and the dotted one is the lower bound for the capacity derived in Ref. \cite{ramsauer2021hopfield}.  \textbf{Bottom}: Gradient descent simulations starting from initial condition $\bxi^1$. Points give values of $\lambda$ where we observe a crossover between  retrieval (higher $\lambda$) and non-retrieval (lower $\lambda$). Solid lines are empirical quadratic fits in  $1/N$. Horizontal lines are the predictions from our $N=+\infty$ theory.}
\label{fig:sph-pd}
\end{figure}

We now consider the stronger requirement that {\it all patterns} can be successfully retrieved. While in the standard Hopfield model and its polynomial generalizations, single and all-patterns retrieval thresholds coincide, in presence of an exponential number of patterns one needs to control exponentially rare events.

In order to derive a lower bound, $\alb\leq \alpha_c$, for the critical value of $\alpha$ such that all patterns are retrieved, we use a union bound: if $p_1$ is the probability of retrieving a typical pattern, the probability $p_c$ that all patterns are retrieved verifies $1-p_c\leq e^{\alpha N}(1-p_1)$. As we have seen, $1-p_1$ is the probability that the free-energy of the REM with $P-1$ energy levels $\ve^\mu=\bxi^{1}\cdot\bxi^{\mu} /N$ (with $\mu\neq 1$) is larger than $\lVert\bxi^{1}\rVert^2/N$. In order to control $p_c$, we thus need to compose the large deviations of the norm of the reference pattern with those of the REM free-energy.

As  proven in Ref. \cite{fedrigo_large_2007} (see also \cite{gardner_probability_1989}), if $\lVert\bxi^{1}\rVert=r\sqrt{N}$ and 
the cumulant generating function is defined for all $\lambda\in  \R$,
differentiable, and strictly convex, then the distribution of the REM free-energy density $\phi$ satisfies a large deviation principle given by $P(\phi)= e^{-N I_{\alpha,r,\lambda}(\phi)}$ with
\begin{equation}
I_{\alpha,r,\lambda}(\phi)=\begin{cases}
        +\infty & \phi<\phi_{\alpha,r}(\lambda)\\
        0 & \phi=\phi_{\alpha,r}(\lambda)\\
        s_{r}(\phi)-\alpha & \phi>\phi_{\alpha,r}(\lambda).
    \end{cases}
\label{eq:I-largedev}
\end{equation}
where $s_r$ is defined in Eq. \eqref{eq-app:s-def}. 
Let us call $\tilde{I}(r)$ instead the rate function for large deviations of $\lVert\bxi^{1}\rVert/\sqrt{N}$. 
Accounting for the fluctuations of $r$ and of the REM free energy conditioned on the value of $r$, it is easy to show that $1-p_1 = e^{-N A(\alpha,\lambda)}$ for large $N$, where 
\begin{align}
	A(\alpha,\lambda) =\inf_{r\in[0,\infty)}\left[\,\tilde{I}(r)+\inf_{\phi\,:\phi>r^{2}}I_{\alpha,r,\lambda}(\phi)\,\right].
	\label{Adef}
\end{align}
Therefore, the union bound provides the lower bound 
\begin{equation}
    \alb(\lambda) = \sup\{\alpha\,:\, A(\alpha,\lambda) -\alpha > 0\}\ .
\label{eq:alb}
\end{equation}
We note that since $\alb\leq\alpha_c\leq\alpha_1$, when $\alb=\alpha_1$ the bound turns into an equality. As we will show, this is always the case for spherical patterns, while for Gaussian patterns there is a gap at any $\lambda$.

\section{Basins of attraction}
\label{sec:Basins}
The full analytic computation of attraction basins requires following a trajectory in time. This is a complicated task, which has not been done in the standard Hopfield model, and which is beyond the reach of our method. 
However, we can obtain a characterization of the size of attraction basins using the energy decomposition of Eq. \eqref{eq:ene_decomp}.
Consider one step of the optimization procedure in Eq. \eqref{eq:update}, starting
from a configuration $\bx$ sampled uniformly at random conditioned on a given norm and angle with the first pattern constraints: $\lVert\bx\rVert = \rho\sqrt{N}$, $\cos(\theta) = \bx \cdot \bxi^1 /\lVert\bx\rVert\lVert\bxi^1 \rVert$.
Then Eq. \eqref{eq:ene_decomp} implies that if 
$\rho\cos(\theta) > \phi_{\alpha,\rho}(\lambda)$
 the first pattern dominates, the gradient is $\bx-\bxi^1$ and the pattern is retrieved after one step of gradient descent at rate $1$. 
On the other hand, if $\rho \cos(\theta) < \phi_{\alpha,\rho}(\lambda)$, the energy is
dominated by the sum over all patterns distinct from $\bxi^1$. Then one needs to study the multi-step dynamics in order to see the final point of gradient descent. 
Therefore, at large $N$, the condition $\rho \cos(\theta) = \phi_{\alpha,\rho}(\lambda)$ identifies the  phase transition. In the region
$\theta < \tc(\alpha,\rho,\lambda)$ 
the pattern is recovered in one step. $\theta_c$ is thus a lower bound to the size of the attraction basin, that we conjecture to be tight.
This phase diagram is valid with high probability for random starting points $\bx$ with given $\rho$ and $\theta$. A more fine-grained analysis of the attraction basis studying all initial conditions (and in particular adverse ones) requires controlling the rare events, similarly to what is done in Sec. \ref{sec:allpatterns}.

Notice that, at large $\lambda$, the value of $\ctc$ is given by geometrical constraints. In the REM analysis, we have $\lim_{\lambda\to{\infty}}\phi_{\alpha,\rho}(\lambda) =\ve_{*}(\alpha,\rho) = \rho\, c_{max}(\alpha)$, where $c_{max}(\alpha)$ is the maximum (in the thermodynamic limit) of 
the $\{\cos(\theta^\mu)\}_{\mu=2}^{e^{\alpha N}}$, where $\theta_\mu$ is the angle between $\bx$ and $\bxi^\mu$: $c_{max}(\alpha)$ gives the angular distance between $\bx$ and the closest neighbor to $\bxi^1$. For discrete-valued patterns, this geometrical analysis is a standard one in error correcting codes, and the corresponding angular distance $c_{max}(\alpha)$ is called the Gilbert-Varshamov distance \cite{Montanari2009InfoPhys,richardson2008modern}. Generalizing its computation to the present case of continuous patterns, we show in Appendix \ref{app:typdist} that $\lim_{\lambda\to\infty} \cos \tc(\alpha,\rho,\lambda)= \rho\, c_{max}(\alpha)$.

It is also interesting to study the basins of attraction when the network is used to store a polynomial number of patterns, $P=c N^k$, much below its capacity, in the spirit of the studies in \cite{Agliari2020}. In this case, we show in Appendix \ref{app:typdist} that for any finite $\lambda$, one has $\cos \theta_c= \sqrt{2\log P/N }\ll 1$. So the attraction basins become very large.

\section{Applications to pattern ensembles}

For {\it spherical patterns}, i.e. with uniform distribution on the sphere $\lVert\bxi\rVert^2=N$,  we compute the auxiliary REM free energy as given by Eq. \eqref{eq:phi_rem-final}.  The threshold $\alpha_1(\lambda)$ for the retrieval of a typical pattern is then identified by $\phi_{\alpha_1, 1}(\lambda)=1$ according to Eq. \eqref{eq:alpha1}. Considering all patterns retrieval instead, evaluation of Eq. \eqref{Adef} yields $A(\alpha,\lambda)=+\infty$ for $\alpha<\alpha_1(\lambda)$, and $A(\alpha,\lambda)=0$ otherwise. Therefore there is a unique retrieval transition line, $\alpha_c=\alpha_1$.
See Appendix \ref{app:spherical} for the details of the calculations.

Fig. \ref{fig:sph-pd} (Top) shows the phase diagram in the $(\alpha,\lambda)$ plane. There is no upper limit to the capacity: at large $\lambda$, one finds $\alpha_c(\lambda)\sim 0.5\log\lambda$, but of course the basins of attraction become very small. Besides the $\alpha_c(\lambda) =\alpha_1(\lambda)$ line, we also report the weaker (but rigorous) lower bound to $\alpha_c(\lambda)$ obtained by extrapolating the results of Ref. \cite{ramsauer2021hopfield} to large $N$ (see Appendix \ref{app:ramsauer}).
 The numerical experiments also presented in Fig. \ref{fig:sph-pd} (Bottom) corroborate the theoretical findings. Because of the exponential capacity, simulations are limited to small values of $N$, but the extrapolation to infinite size is nevertheless in reasonable agreement with our theory. Additional numerical results are presented in Appendix \ref{app:numerical}.

In Fig. \ref{fig:basin_spherical} we show the critical size for the basins as discussed in Section \ref{sec:Basins}. Considering a typical pattern, recovery is possible from a random initial condition on the sphere at an angle $\theta$ from it as long as $\theta<\theta_c(\alpha, \lambda)$. The expression for $\theta_c$ is given in Appendix \ref{app:spherical}. We see that the basin size decreases monotonically with $\alpha$ as expected. For a given $\alpha$ and increasing $\lambda$, we find 3 regimes: 1) No retrieval for $\lambda<\lambda_1(\alpha)$; 2) Monotonically increasing basin size for $\lambda_1(\alpha) <\lambda < \lambda_*(\alpha)$; 3) Basin size frozen to $\theta_*(\alpha)$ for $\lambda > \lambda_*(\alpha)$. 
The value $\theta_*(\alpha)$ corresponds to the typical distance of nearest patterns. The REM is in a condensed phase in this last regime.

\begin{figure}[t]
	\includegraphics[width=\columnwidth, trim = {0 0 0 28}, clip]{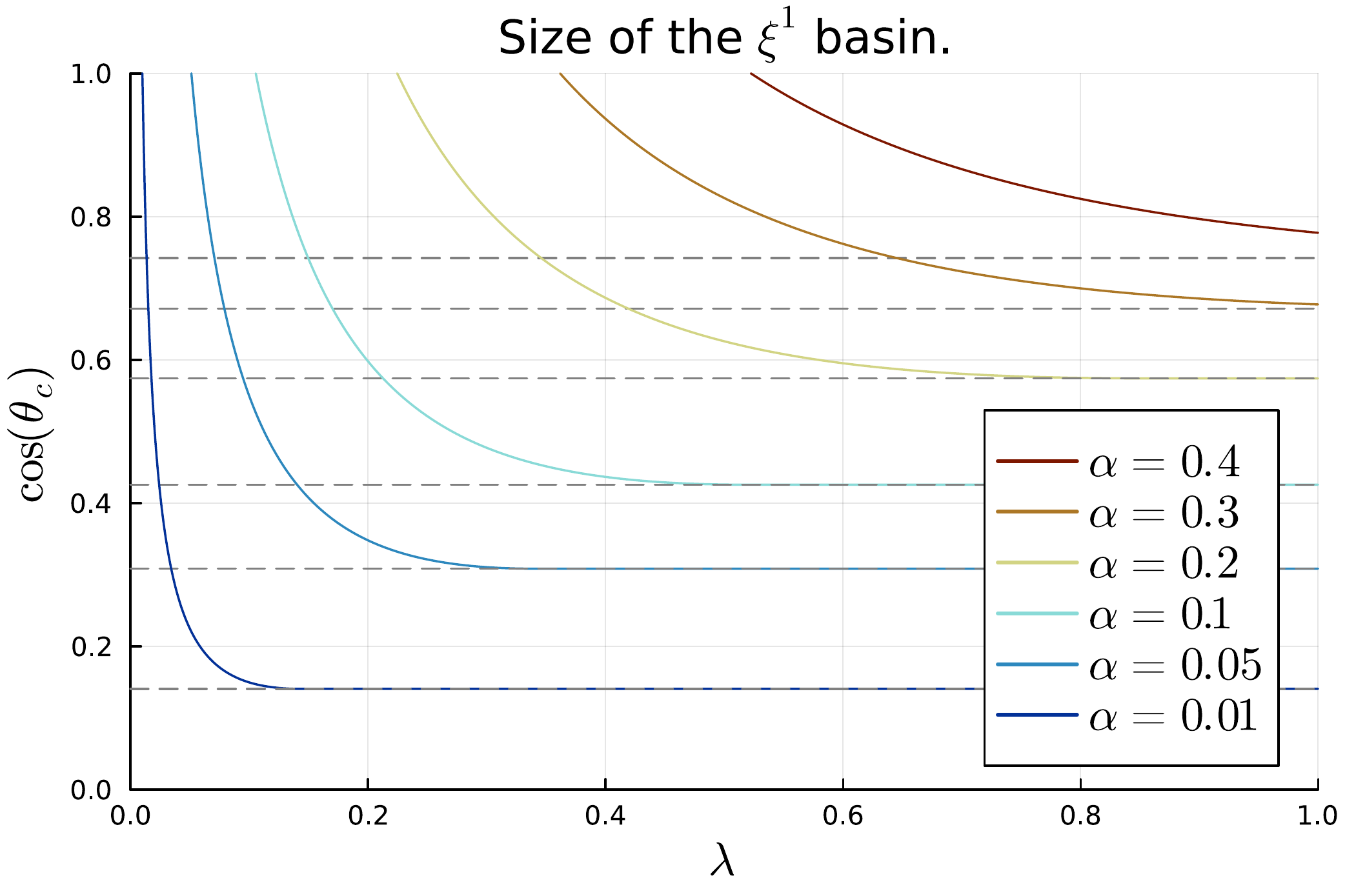}
	\caption{Characterization of the basins of attraction given by the maximum angle $\theta_c(\alpha,\lambda)$ between a typical pattern and a random initialization such that the pattern is retrieved with high probability. The patterns here follow a spherical distribution and the configuration is also initialized on the hypersphere. Horizontal lines correspond to the angle of the nearest pattern.} 
	\label{fig:basin_spherical}
\end{figure}

In the case of {\it Gaussian patterns}, fluctuations in the norm of the patterns lead to larger deviations from the typical behavior compared to the spherical case. 
The phase diagram derived in Appendix \ref{app:gaussian} and summarized in Fig. \ref{Fig-gauss-pd} thus differs significantly from the one of spherical patterns. The capacity threshold for typical pattern retrieval, $\alpha_1(\lambda)$,  saturates at $\lambda=1$ to the value $\alpha_1=1/2$. The lower bound to $\alpha_c$ no longer coincides with $\alpha_1$. It saturates at large $\lambda$ to the value $\alb= \log2\,/\,4$. Finally, in the region $\alpha>\alpha_1$ where the patterns are not retrieved, the noise term dominating the energy, which is given by the REM term, can be either in a condensed phase 
or not 
depending on the value of $\lambda$.

Fig. \ref{fig:pairwise_comparison} of Appendix \ref{app:numerical} shows that the numerical estimate of $\alpha_c$ is close to our analytic lower bound.

\begin{figure}[t]
	\centering    
	\includegraphics[width=\columnwidth, trim={5 0 0 25}, clip]{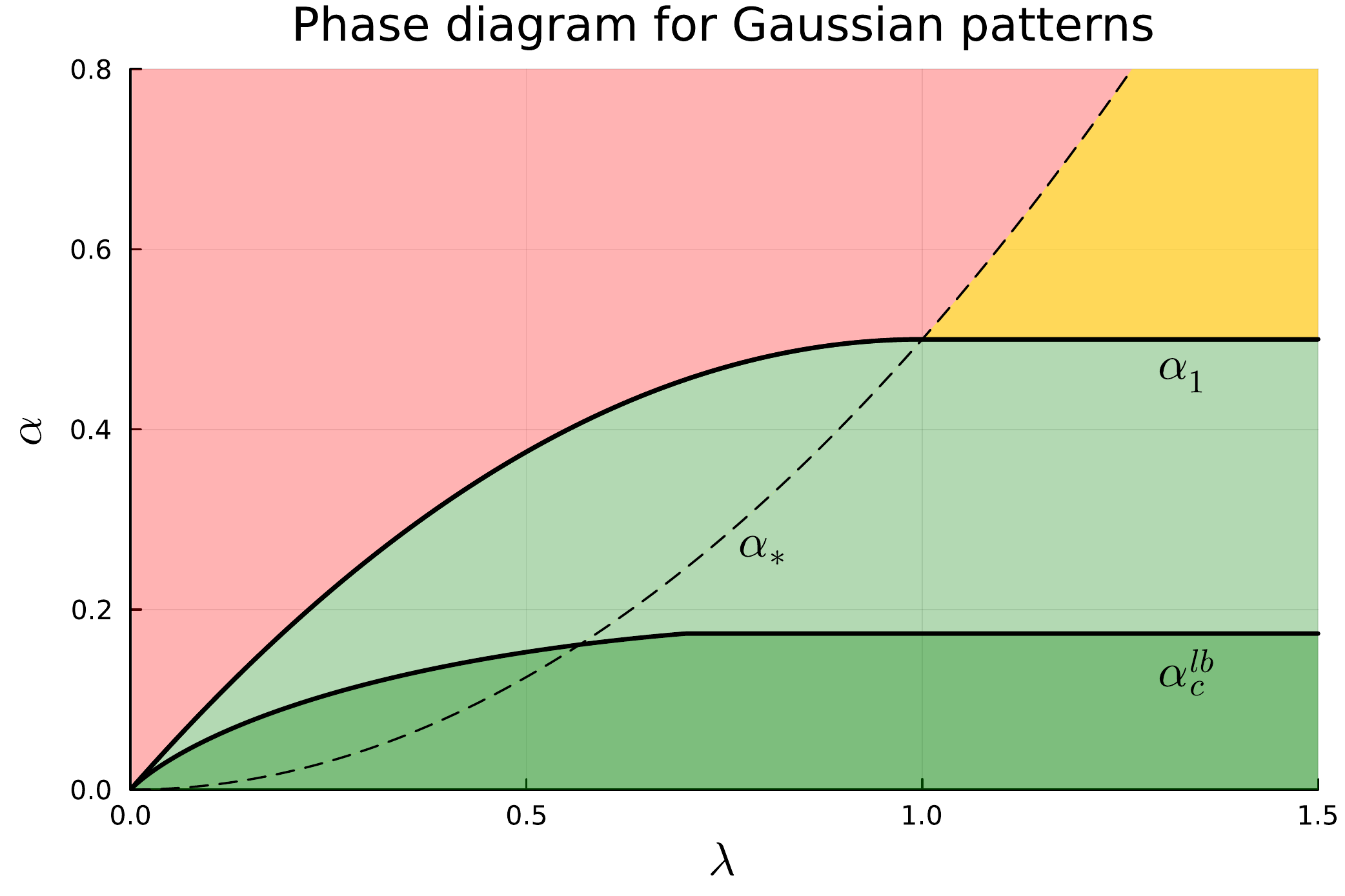}
	\caption{Phase diagram for Gaussian patterns, showing the critical capacity for retrieval of a typical pattern, $\alpha_1$, as well as the lower bound for the capacity of retrieval of all patterns, $\alb$. The $\alpha_*$ line is the condensation transition in the auxiliary REM. In the non-retrieval regime $\alpha>\alpha_1$ there exist two phases, one with $\alpha>\alpha_*$ where the energy is dominated by an exponential number of patterns, and the condensed phase $\alpha_1<\alpha<\alpha_*$.}
	\label{Fig-gauss-pd}
\end{figure}

\section{Scaled dot-product}
\label{sec:scaled}
The dot-product found in the softmax operation of Transformer architectures is commonly scaled by a $\sqrt{N}$ factor \cite{vaswani2017attention} so that totally uncorrelated keys and queries give an $O(1)$ exponent.  Therefore we consider a scaling regime where $\lambda=\tilde{\lambda} / N^a$ for some $a\in (0,1)$. The energy function now reads
\begin{equation}
	E(\bx) = -\frac{N^a}{\tilde\lambda}\log \sum_\mu e^{\frac{\tilde\lambda}{N^a} \bx \cdot \bxi^\mu} + \frac{1}{2}\lVert \bx\rVert^2.
\end{equation}
The interesting regime is now when the number of patterns scales as $P = \exp(\tilde{\alpha} N^{1-a})$ so that we have a competition between the noise contribution $\Phi$ given by the REM and the signal term $\frac{\tilde\lambda}{N^a} \rVert\bxi^1\lVert^2\approx \tilde{\lambda} N^{1-a}$ in Eq. \eqref{eq:ene_decomp} when $\bx=\bxi^1$. Then, $\lambda N\Phi =\tilde{\alpha} N^{1-a}$ for any pattern ensemble.
The typical pattern retrieval threshold in this setting always coincides with the all-patterns retrieval one, and they take the simple value $\tilde{\alpha}_1=\tilde{\alpha}_c=\tilde{\lambda}$. In Appendix \ref{app:scaled} we present numerical experiments supporting this result.

\section{Conclusions}
Dense associative memories have exponential capacity, at the cost of using an energy function that takes an exponential number of operations. This gives a completely new regime where we can use statistical physics to determine the asymptotic memory capacity of large networks, as well as their attraction basins for zero-temperature dynamics starting from random configurations. Generalizing these results to properties of the free-energy landscape (at finite temperature), or to worst-case initial conditions is an interesting challenge: subtle effects due to the exponential number of patterns, and resulting rare events, are to be taken into account. On the other hand, it should be possible to establish rigorously our zero-temperature results and to extend them to other factorized pattern distributions, such as the case of binary patterns and continuous neurons. Finally, it would be interesting to extend the study to patterns generated from a hidden-manifold \cite{negri2023hiddenmanifold}, and to
explore implications for Transformer architectures.
C.L. acknowledges funding from the European Union – Next Generation EU (MIUR PRIN 2022 and MIUR PRIN PNRR 2022).
\bibliographystyle{unsrturl}
\bibliography{bibliography}

\onecolumngrid
\appendix

\section{Computation of the noise function $\Phi(x)$}
\label{app:noisefunction}
This Appendix aims to present in a self-contained way the Random Energy Model (REM)
analysis
which is used to derive the noise function $\Phi(\bx)$. This analysis
uses the methods that were initially developed by Derrida
\cite{Derrida1981REM}, and generalizes them to a broader range of problems,
including the ones which are encountered when computing
$\Phi(\bx)$. In the more than 40 years since Derrida's work, numerous papers have appeared on the
REM, its mathematical formalism \cite{ruelle1987mathematical,bovier2002fluctuations}, and its generalizations \cite{bouchaud1997universality}. The present Appendix
cannot give a fair account of all these developments, therefore it will present the main ideas without going into the details of mathematical
proofs, but we hope that it will give a reasonably self-contained set
of ideas and methods for colleagues who are not experts in
statistical physics of disordered systems. The Chapter 5 of 
\cite{Montanari2009InfoPhys} also provides a self-contained
introduction to the REM itself. We mainly follow here the line of approach
of that chapter, extending it to the more general setup useful for our
present problem. This Appendix is divided into four subsections. Section \ref{app:rem1}    
 gives the general analysis of partition functions of a REM type, when the statistics of the random variables is known. Section \ref{app:rem2} studies the statistics of the random variables
that appear in the noise $\Phi(\bx)$ in dense
associative memories. Section \ref{app:rem3-summary} summarizes the various steps of the
derivation of the noise function. Section \ref{app:rem4} presents an alternative approach to the computation of $\Phi(\bx)$
using replicas: it gives back the same results as the direct approach
of the first three subsections, in a more compact way. Although this
is neither particularly transparent nor rigorous, we include it here
because it is an interesting
approach in itself and it also gives a path towards studying large deviations.

\subsection{REM partition functions}
\label{app:rem1}
Consider a set $\cS$ of $P=e^{\alpha N}$ independent random variables $\varepsilon^\mu$
which are i.i.d.  random variables with probability density function (pdf) $p_N(\varepsilon)$.  The pdf is assumed to satisfy, at large $N$, a large deviation
principle with a certain rate function $s(\ve)$.
That is, for large $N$ and any $a<b$ we have, at the leading exponential order we have
\begin{align}
\int_a^b d\ve \; p_N(\varepsilon)\approx e^{-N \max_{\ve \in [a,b]} s(\ve)}.
\label{eq:rate-s}
\end{align}
A classical choice \cite{Derrida1981REM} for the pdf is $p_N = \mathcal{N}(0, 1/N)$.
In the cases that we shall study, the function $s(\ve)$ is a
convex non-negative function, which vanishes at the typical value of the random
variable. In the usual REM, the independent random variables $
\ve^\mu$ are called energies, hence the name of the model. The energies in the auxiliary REM model are not to be confused with
the main object of our study,
the energy function $E(\bx)$ of the dense associative memory in Eq. \eqref{eq:enedef}.
The REM  partition function is defined as
\begin{align}
  Z=\sum_{\mu=1}^P e^{\lambda N \ve^\mu}\ .
  \label{REMdef}
\end{align}
Notice that at odds with the usual statistical physics definition, the energies appear in the exponent with a plus sign instead of a minus.
It depends on a parameter $\lambda$ that we choose
positive without loss of generality. It plays the role of an inverse temperature: increasing $\lambda$, the $\ve^\mu$ with higher values will give the dominant contribution to $Z$. 
One can define the free-energy density of the REM as $\Phi=\frac{1}{\lambda
  N}\log Z$.
A main consequence of the independence of the variables in $\cS$, and of the specific
large-deviation form of their distribution is that, in the large $N$
limit, the random variable $\Phi$ concentrates: its distribution
becomes peaked around its typical value
\begin{equation}
    \phi_\alpha(\lambda) = \lim_{N\to\infty} \mathbb{E}\,\Phi,
\end{equation}
with a standard deviation that goes to zero as $1/\sqrt{N}$. Let us see how one can
compute the typical value of the
free energy density in the large $N$ limit, $\phi$,
which depends on $\alpha,\lambda$ and on the rate function $s(\ve)$, and justify the concentration property.
In the large $N$ limit, let us call $\mathcal{N}_{[a,b]}$ the number of random variables of $\cS$ (among the $P=e^{\alpha N}$ it contains) which
are in the interval $ [a,b] $. Its expected value is 
\begin{align}
\mathbb{E}\, \mathcal{N}_{[a,b]}= e^{N(\alpha - \max_{\ve \in [a,b]}
  s(\ve))},
\end{align}
therefore the average density of random variables around $\ve$ is $e^{N(\alpha -
  s(\ve))}  $. 
The function $\alpha -  s(\ve)$ is a concave
function of $\ve$, it vanishes for certain values $\ve=\ve_m(\alpha)$ and $\ve=\ve_*(\alpha)$, with
$\ve_m(\alpha)<\ve_0 <\ve_*(\alpha)$, it is positive for $\ve \in [\ve_m(\alpha),\,\ve_*(\alpha)]$ and it is
negative outside of this interval.
Using the first- and second-moment methods, one can prove that at large
$N$, $\frac{1}{N}\log 
\mathcal{N}_{[a,b]}$ concentrates around $\max_{\ve \in
  [a,b]}\psi_\alpha(\ve)$, where:
\begin{equation}
\psi_\alpha(\ve) =
\begin{cases}
\alpha -  s(\ve) & \text{if } \ve \in [\ve_m(\alpha),\,\ve_*(\alpha)]\\
-\infty  & \text{otherwise}
\end{cases}
\label{eq:psityp}
\end{equation}

We shall not detail this proof, referring the reader to \cite{Derrida1981REM}. Let us just mention the ideas behind
the proofs. The first
moment method uses Jensen's inequality $ \log 
\mathbb{E}\mathcal{N}_{[a,b]}  \geq \mathbb{E} \log 
\mathcal{N}_{[a,b]}$ in order to show that, when
$\alpha -  s(\ve) <0$, the average number of variables in $\cS$ around
$\ve$ is exponentially small in $N$, which implies that the typical
number of energies is zero. This explains the $-\infty$ case in Eq. \eqref{eq:psityp}. The second
moment method uses the independence of the variables to show that,
when $ \mathbb{E} \mathcal{N}_{[a,b]}$ is exponentially large in $N$, the relative fluctuations
$ \sqrt{\mathbb{E} (\mathcal{N}_{[a,b]}^2)-(\mathbb{E} \mathcal{N}_{[a,b]} )^2}/ (\mathbb{E} \mathcal{N}_{[a,b]})$
are exponentially small, leading to the concentration result \eqref{eq:psityp}.

Let us study the partition function \eqref{REMdef}. Using the
concentration property for the density of levels
\eqref{eq:psityp}, one obtains the large $N$ behaviour
\begin{align}
Z\approx\int_{\ve_m(\alpha)}^{\ve_*(\alpha)}d\ve \; e^{N (\alpha
  -s(\ve)+\lambda \ve)}.
  \end{align}
This integral can be evaluated using Laplace's method which gives
\begin{align}
\lim_{N\to \infty} \frac{1}{N}\log Z= \max_{\ve \in [\ve_m(\alpha),\,\ve_*(\alpha)]} \ \alpha
  -s(\ve)+\lambda \ve,
\end{align}
where equality is in the sense of almost sure convergence.
Depending on the value of $\lambda>0$, we find two regimes separated by
a critical value $\lambda_*(\alpha)=s'(\ve_*(\alpha))$. If $\lambda<\lambda_*(\alpha)$,
the maximum is found at a value $\ve=\tilde{\ve}(\lambda)$, with $\ve_m(\alpha)<\tilde{\ve}
<\ve_*(\alpha)$, obtained as the stationary point of $\alpha
  -s(\ve)+\lambda \ve$.
In this regime, an exponentially large
number of variables $\ve^\mu$ contribute to the partition function (the
entropy of the REM is proportional to $N$). On the contrary when
$\lambda\geq\lambda_*(\alpha)$, the maximum is at $\ve_*(\alpha)$ and the entropy vanishes. The transition between these two regimes is called condensation transition.

This discussion leads to the following result:   the free-energy
density of a REM-type model, $\frac{1}{\lambda
  N}\log Z$, concentrates at large $N$ to the value
\begin{align}
  \phi_\alpha(\lambda)&= 
  \begin{cases}
   \tilde{\ve}(\lambda) + \frac{\alpha
  -s(\tilde{\ve}(\lambda))}{\lambda}& 
        \lambda<\lambda_*(\alpha) \\
\ve_*(\alpha) &
    \lambda\geq\lambda_*(\alpha)
\end{cases}
    \label{eq:phi_rem-gen}
\end{align}
where $\tilde{\ve}(\lambda)$ is defined by $\lambda=s'(\tilde{\ve})$, 
The critical value $\lambda_*(\alpha)$ instead is found  when $\ve(\lambda)$ becomes
equal to the maximal value $\ve_*(\alpha)$ of the $P=e^{\alpha N}$ random
variables: $\lambda_*(\alpha)=s'(\ve_*(\alpha))$.

In the next sections, we discuss how to compute the rate function $s(\ve)$ and summarize the steps needed to apply the formalism to our original dense associative memory problem.

\subsection{Determination of the rate function}
\label{app:rem2}
Let us go back to the definition  $\eqref{eq:def-Phi}$ of the noise term in our analysis of
the dense associative memory in presence of $P=e^{\alpha N}$ memories $\bxi^\mu \in \mathbb{R}^N$ independently sampled from a density function $p_N(\bxi)$:
\begin{align}
\Phi(\bx)=\frac{1}{N\lambda}\log \left(\sum_{\mu=2}^P e^{\lambda \bx\cdot\bxi^\mu}\right).
\end{align}
We consider the vector $\bx$, chosen independently from patterns
$\{\bxi^\mu\}_{\mu=2,...,P}$. We are in a situation similar to the REM partition
function of Eq. $\eqref{REMdef}$, where the i.i.d.  random variables are
$\ve^\mu= \frac{1}{N}\bx\cdot\bxi^\mu$.
We need to establish that the large deviation principle $\eqref{eq:rate-s}$ applies and compute the rate function $s(\ve)$.

We introduce the generating function:
\begin{align}
\Gamma(\lambda, \bx)=\int d\bxi p_N(\bxi)\,e^{\lambda  \bx\cdot\bxi}
\end{align}
Where the expectation is over the pattern distribution $p_N\bxi)$.
A priori, this generating function depends on the full vector
$\bx$. Assuming that the distribution  $p_N(\bxi)$ is rotationally
invariant, one finds that $\zeta_{\bx}$ depends on $\bx$ only through
its norm, $\bx=\rho \sqrt{N}$. We also assume that $\Gamma$ satisfies a large deviation principle with a (negative) rate function $\zeta_\rho(\lambda)$ , that is for large $N$ we have
\begin{align}
	\Gamma(\lambda,\bx)\approx e^{N\zeta_\rho(\lambda)}\
  \label{zeta-rho-def}
\end{align}
Under the assumptions of G{\"a}rtner-Ellis theorem, 
this implies that the  distribution $\rho_N(\ve)$ satisfies a large
deviation principle with a rate function $s_\rho(\ve)$, which is
the Legendre transform of $\zeta_\rho(\lambda)$. In fact, writing
\begin{align}
\Gamma(u,\bx)\approx\int d\ve\ e^{-Ns_\rho(\ve)}e^{\lambda N \ve}\approx e^{N\zeta_\rho(u)},
\end{align}
we find the relations 
\begin{align}
 \zeta_{\rho}(\lambda)= \sup_\ve\ \lambda \ve-s_\rho(\ve)
 	\label{eq:zeta-rho}
\end{align}
and
\begin{align}
  s_\rho(\ve)= \sup_\lambda \  \lambda\ve- \zeta_\rho(\lambda).
  \label{eq-app:s-def}
\end{align}

\begin{figure}[!htb]
	\includegraphics[width=0.6\textwidth]{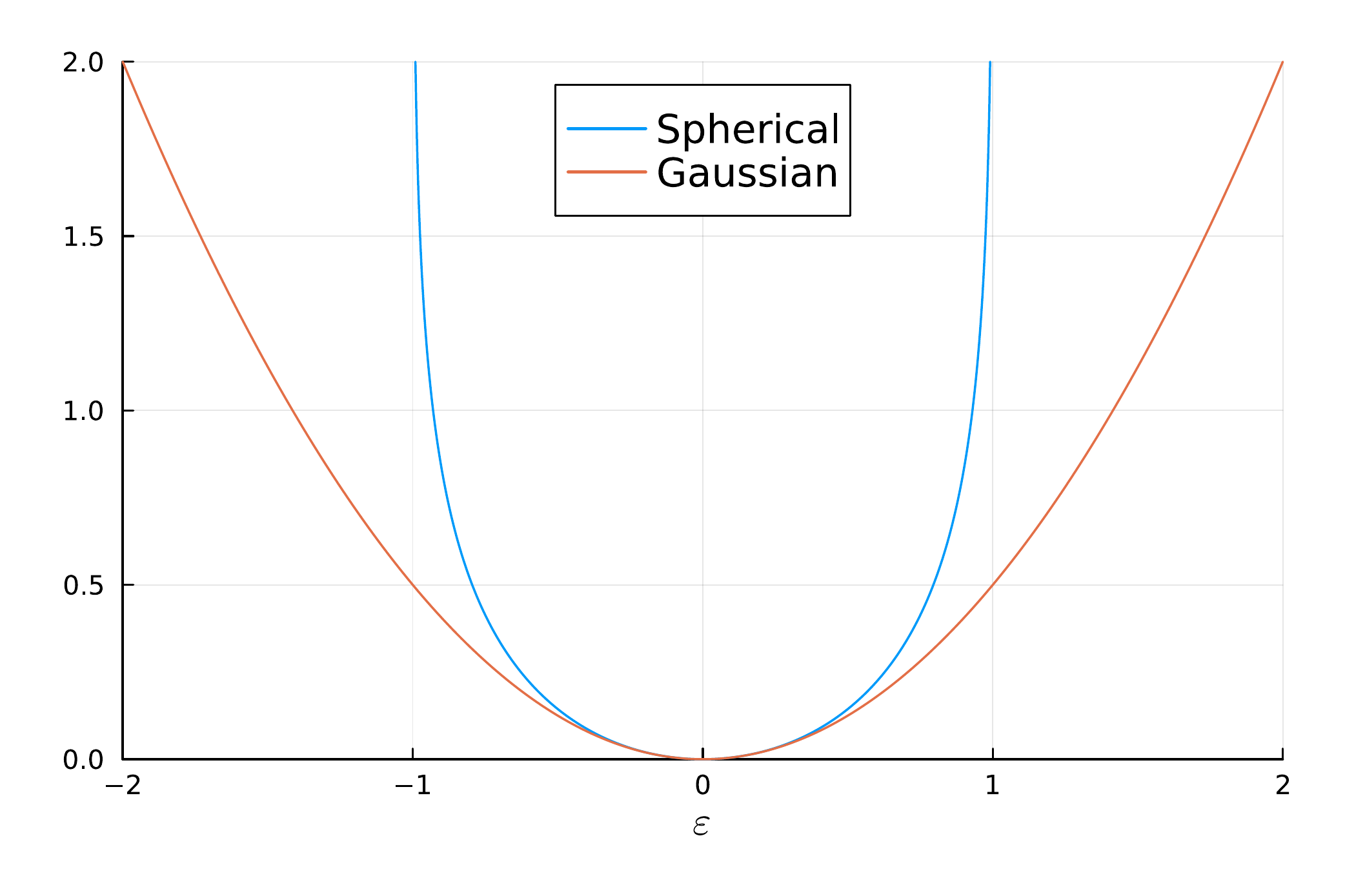}
	\caption{The REM rate function $s_\rho(\ve)$ for $\rho=1$ and the two  pattern ensembles considered in this paper.}
	\label{fig:ratefunction}
\end{figure}

The rate function $s_\rho(\ve)$ is shown in Fig. \ref{fig:ratefunction} for the Gaussian and Spherical pattern ensembles. 

\subsection{Summary of the noise function computation}
\label{app:rem3-summary}
We summarize here the general approach for finding the large $N$ limit of the REM-like noise term $\Phi(\bx)$ in \eqref{eq:def-Phi}, with $P=e^{\alpha N}$ patterns independently sampled from  a
rotationally invariant distribution $ p_N(\bxi)$.

For increasing $N$, we consider a sequence of vectors  $\bx\in \mathbb{R}^N$ and patterns $\bxi^\mu \in \mathbb{R}^N$, such that $\bx$ is chosen independently from $\{\bxi^\mu\}_{\mu=2,...,P}$ and with norm $\lVert \bx \rVert=\rho \sqrt{N}$. Thanks to rotational invariance and to concentration of the REM free energy on its expected value (the self-averaging property in statistical physics jargon), we have

\begin{equation}
	\phi_{\alpha,\rho}(\lambda) = \lim_{N\to\infty} \frac{1}{N} \Phi(\bx).
\end{equation}

The asymptotic value $\phi_\rho(\lambda)$
is thus obtained as follows:
\begin{enumerate}
\item Find the generating function
  $\zeta_\rho(\lambda)$ defined in \eqref{zeta-rho-def}:
  \begin{equation}
  	\zeta_\rho(\lambda) = \lim_{N\to+\infty} \frac{1}{N} \log\mathbb{E}_{\bxi}\,e^{\lambda \bx \cdot \bxi}.
  	\label{eq:zeta-rho-final}
  	 \end{equation}
\item
Using the Legendre transform \eqref{eq-app:s-def}, find the rate function
$s_\rho(\ve) =\sup_\lambda \  \lambda\ve- \zeta_\rho(\lambda)$.
\item
  For a given value of $\alpha$, find the
values $\ve_m(\alpha, \rho) <\ve_*(\alpha, \rho)$ which are the two solutions of $\alpha
  -s_\rho(\ve) = 0$. We are assuming here  $s_\rho(\ve)$ to be concave and unbounded above. For bounded $s_\rho(\ve)$, the argument has to be slightly modified.
\item Obtain the condensation threshold  $\lambda_*(\alpha,\rho) = s_{\rho}^\prime(\ve_*(\alpha, \rho))$.

\item Obtain the energy level $\tilde\ve(\lambda, \rho)$ dominating the uncondensed phase as the stationary point of $\lambda \ve - s_\rho(\ve)$.
\item
  Finally compute
  $\phi_{\alpha,\rho}(\lambda)$ as
 
  \begin{align}
  	\phi_{\alpha,\rho}(\lambda)= 
  	\begin{cases}
  		\tilde{\ve}(\lambda, \rho) + \frac{\alpha
  			-s_\rho(\tilde{\ve}(\lambda, \rho))}{\lambda}=\frac{\alpha
  			+\zeta_\rho(\lambda)}{\lambda}& 
  		\lambda<\lambda_*(\alpha, \rho) \\
  		\ve_*(\alpha, \rho) &
  		\lambda\geq\lambda_*(\alpha, \rho).
  	\end{cases}
  		\label{eq:phi_rem-final}
  	\end{align}
  	
  \end{enumerate}
\subsection{An alternative approach: replicas}
\label{app:rem4}
We now show how to obtain the same result for the free energy using the replica method, generalizing the derivation in Refs. \cite{Derrida1981REM, Montanari2009InfoPhys}. It is convenient to use rotational invariance and rewrite the free energy as
\begin{equation}
	\phi_{\alpha,\rho}(\lambda)=\lim_{N\to\infty}\frac{1}{\lambda N}\mathbb{E}\log\sum_{\mu=2}^P e^{\lambda\rho\sqrt{N}\xi_{1}^{\mu}}.
\end{equation}
It is clear that the $\rho$ dependence could be reabsorbed in the $\lambda$ one (or vice-versa), but we will keep both parameters for consistency with the main discussion.
We then consider an integer number $n>0$ of replicas, with $n$ to be sent to zero by analytical continuation at the end of the computation. The average replicated partition function reads
\begin{align}
	\mathbb{E}\,Z^{n} & =\mathbb{E}\sum_{\mu_{1},\dots,\mu_{n}}e^{\lambda\rho\sqrt{N}\sum_{a=1}^n\xi_{1}^{\mu_{a}}}.
\end{align}

In order to decouple the average over patterns, we introduce the variables $m_\mu$ defined by $m_{\mu}=\sum_{a=1}^n\delta_{\mu_{a},\mu}$ for a given choice of the labels $\mu_a$. Defining the combinatorial factor
\begin{equation}
	M(\{m_{\mu}\}_{\mu})=\sum_{\{\mu_{a}\}_{a}}\prod_{\mu}\delta(m_{\mu}-\sum_{a}\delta_{\mu_{a},\mu}),
\end{equation}
we obtain
\begin{align}
	\mathbb{E}\,Z^{n} & =\mathbb{E}\ensuremath{\sum_{\{n_{\mu}\}_{\mu}}M(\{m_{\mu}\}_{\mu})\ e^{\sum_{\mu}m_{\mu}\rho\sqrt{N}\xi_{1}^{\mu}}}\\
	& =\sum_{\{m_{\mu}\}_{\mu}}M(\{m_{\mu}\}_{\mu})\ e^{N\sum_{\mu}\zeta_{\rho}(\lambda m_{\mu})},
\end{align}
where $\zeta_{\rho}(\lambda)$ is the generating function given by
the single pattern expectation
\begin{equation}
	\zeta_{\rho}(\lambda)=\lim_{N\to\infty}\frac{1}{N}\log\mathbb{E}_{\bxi}\,e^{\sqrt{N}\lambda\rho\xi_{1}}.
\end{equation}
Notice that this is the same $\zeta_{\rho}(\lambda)$ of Eqs. \eqref{eq:zeta-rho} and \eqref{eq:zeta-rho-final}.
We will now make an ansatz corresponding to one step of replica symmetry breaking (1RSB) \cite{mezard1987spin, Montanari2009InfoPhys} for the values of $m_{\mu}$ dominating the summation. 
\newline
\newline
\textbf{1RSB ansatz: }\emph{The $n$ replicas $\mu_{a}$ are divided
into $n/m$ groups of size $m$.  Within a group,
all $\mu_{a}$ are the same. Different groups have different $\mu_{a}$.
Therefore $m_{\mu}=m$ for $\frac{n}{m}$ of the patterns, and $m_{\mu}=0$
for the others.}
\newline
\newline
The parameter $m$, which has to be optimized, is called the Parisi 1RSB parameter. Within this ansatz, we have $M=e^{n\frac{\alpha}{m}N}$, from which one obtains

\begin{equation}
	\phi_{\alpha,\rho}(\lambda)=\frac{1}{\lambda}\,\inf_{m\in[0,1]}\ \left[\frac{\alpha}{m}+\frac{1}{m}\zeta_\rho(\lambda m)\right] \ .
 \label{eq-app:phi}
\end{equation}

For any given $\rho$, the optimal value for the Parisi parameter $m$
is increasing in $\alpha$. Moreover, we also assume the monotonicity of $\zeta_\rho(\lambda)$, which implies that the optimal $m$ is monotonous in $\lambda$ as well. We now define as $m_*(\alpha,\rho, \lambda)$ the value of $m$ obtained imposing a stationarity condition in Eq. \eqref{eq-app:phi}:

\begin{equation}
	-\frac{\alpha}{m_*^{2}}-\frac{1}{m_*^{2}}\zeta_\rho(\lambda m_*)+\frac{\lambda}{m_*}\zeta_\rho^{\prime}(\lambda m_*)=0
 \label{eq:stationary_m}
\end{equation}
The condensation threshold $\lambda_{*}(\alpha, \rho)$ is defined by
the value of $\lambda$ for which the stationary $m$ is exactly one:
\begin{equation}
	m_{*}(\alpha, \rho, \lambda_*)=1.
\end{equation}
Solving the same equation for $\alpha$ instead yields the equivalent parametrization of the critical surface $\alpha_*(\rho, \lambda)$.
The optimal value of $m$ constrained to the interval $[0,1]$ in Eq. \eqref{eq-app:phi} is then given by 

\begin{eqnarray}
	m(\alpha,\rho, \lambda) & = & \begin{cases}
		1 & \lambda <\lambda_{*}(\alpha, \rho)\\
		m_{*}(\alpha,\rho, \lambda) & \lambda\geq\lambda_{*}(\alpha, \rho).
	\end{cases}
\label{eq-app:m-optimal}
\end{eqnarray}
The two phases at $\lambda <\lambda_{*}$ and $\lambda >\lambda_{*}$ are called dynamical and static 1RSB phase respectively.
 The final result is given by Eq. \eqref{eq-app:phi}, to be evaluated at the optimal value of $m$ given in Eq. \eqref{eq-app:m-optimal}. 

The equivalence between the free energy expression in Eq. \eqref{eq-app:phi} and the one given in Eq. \eqref{eq:phi_rem-final}  is not immediately obvious for $\lambda <\lambda_*$ and it is due to the general Legendre structure of the 1RSB formalism. Considering the free energy functional as function of $m$, in our case $\phi(m) = \frac{1}{\lambda}\,(\frac{\alpha}{m}+\frac{1}{m}\zeta_\rho(\lambda m))$, one can in fact show  \cite{Montanari2009InfoPhys} that it can be decomposed in a complexity and energy contribution, $\phi(m) = \frac{1}{\lambda m} (\Sigma(\varepsilon(m))+ \lambda m\, \varepsilon(m)$). Since $\frac{\partial\phi}{\partial \varepsilon}=0$ due to the stationarity of the action,  we have $\frac{\partial\phi}{\partial m}=-\frac{1}{\lambda m^2} \Sigma(\varepsilon(m))$. Therefore, Eq. $\eqref{eq:stationary_m}$ selects the value $m_*$ which gives zero complexity and the maximum allowed energy value $\varepsilon_*$. This completes the replica description of the average free energy of the REM.

The replica analysis can be extended to the computation of the large deviation function. This is done considering the partition function at finite  integer $n$, and using a 1RSB ansatz and analytic continuation to obtain
\begin{equation}
G_{\alpha,\rho,\lambda}(n) = \lim_{N\to\infty}\frac{1}{N}\log \mathbb{E}\,Z^n.
\end{equation}
for generic $n$.
Thanks to the G\"{a}rtner-Ellis theorem, one can then perform a
Legendre transform and obtain the convex-hull
$\hat{G}_{\alpha,\rho,\lambda}(\phi)$ of the true rate function
$I_{\alpha,\rho,\lambda}(\phi)$. Unfortunately, in the uncondensed
phase $\lambda < \lambda_*$, the convex-hull $\hat{G}$ differs from
the expression for $I_{\alpha,\rho,\lambda}$ proved by in
Ref. \cite{fedrigo_large_2007} and reported in
Eq. \eqref{eq:I-largedev}. For a thorough analysis of the large
deviations in the REM as computed by the replica method, we refer the
reader to Chap. 2 in Ref. \cite{pastore_replicas_2021}.

\section{Typical distances}
\label{app:typdist}
The typical distance between a configuration and the nearest patterns is related to the attraction basin size at large $\lambda$ and can be studied as follows, generalizing to the case of continuous variables the 'Gilbert-Varshamov' derivation in error-correcting codes \cite{Montanari2009InfoPhys,richardson2008modern}. 
We take a configuration $\bx$ with $\lVert \bx\rVert=\sqrt{N}$ as the reference configuration, 
since as argued in Section \ref{sec:Basins} it is sufficient to consider the case $\rho=1$. We could also take $\bx =\bxi^1$, the only requirement is that  $\bx$ is chosen independently from the $\mu\geq2$ patterns. We study the overlaps $c^\mu=\bx \cdot\bxi^\mu/N$. Conditioning on $\bx$, the $c^\mu$ are independent random variables. For a given realization of the patterns, let us call $\mathrm{N}(c)$ the number of patterns with overlap $c^\mu=c$.
We first compute the annealed average
\begin{align}
   \mathbb{E}_{\bxi^{2:P}}\,\mathrm{N}(c) &= \mathbb{E}_{\bxi^{2:P}}\sum_{\mu\geq 2} \delta(Nc - \bx\cdot\bxi^\mu) \\
    &=\mathbb{E}_{\bxi^{2:P}}\sum_{\mu=2}^P \int \frac{dv}{2\pi/N}\,\ e^{-N vc+v\bxi^1\cdot\bxi^\mu} \\
   &=\int \frac{dv}{2\pi/N}\;e^{N(\alpha -vc+ \zeta_{1}(v))}
\end{align}
 where the
integral over $v$ is along the imaginary axis. Using the saddle point
method we obtain 
\begin{equation}
    \mathrm{S}^{ann}_\alpha(c)=\lim_{N\to\infty}\frac{1}{N}\log
\mathbb{E}_{\bxi^{2:P}}\mathrm{N}(c)= \st_v\ \left[\alpha -v c+\zeta_{1}(v)\right] = \alpha -s_{1}(c),
\end{equation} 
where $\st$ denotes the stationary point (a minimum in this case).
Because of the independence of the random variables $c^\mu$, the usual REM argument shows that the random variable $\frac{1}{N}\log
\mathrm{N}(c)$ converges almost surely to a deterministic quantity:
\begin{equation}
S_\alpha(c)=\lim_{N\to\infty}\frac{1}{N}\log
\mathrm{N}(c)=
\begin{cases}
    \mathrm{S}^{ann}_\alpha(c) & \mathrm{S}^{ann}_\alpha(c)\geq 0\\
    -\infty &  \mathrm{S}^{ann}_\alpha(c)< 0.
\end{cases}
\end{equation}
Therefore $c_{max}(\alpha)$ is obtained as the largest root of $\mathrm{S}_\alpha^{ann}(c_{max})=0$. Notice that this is exactly the condition $\alpha =s_{1}(\ve_*)$ determining $\ve_*(\alpha,\rho=1)$. In fact, the largest physical (non-negative complexity) energy level corresponds to the maximum overlap $c_{max}(\alpha)=\ve_*(\alpha,\rho=1)$.

\begin{figure}[t]
	\includegraphics[width=0.7\textwidth, trim={10 0 0 10}, clip]{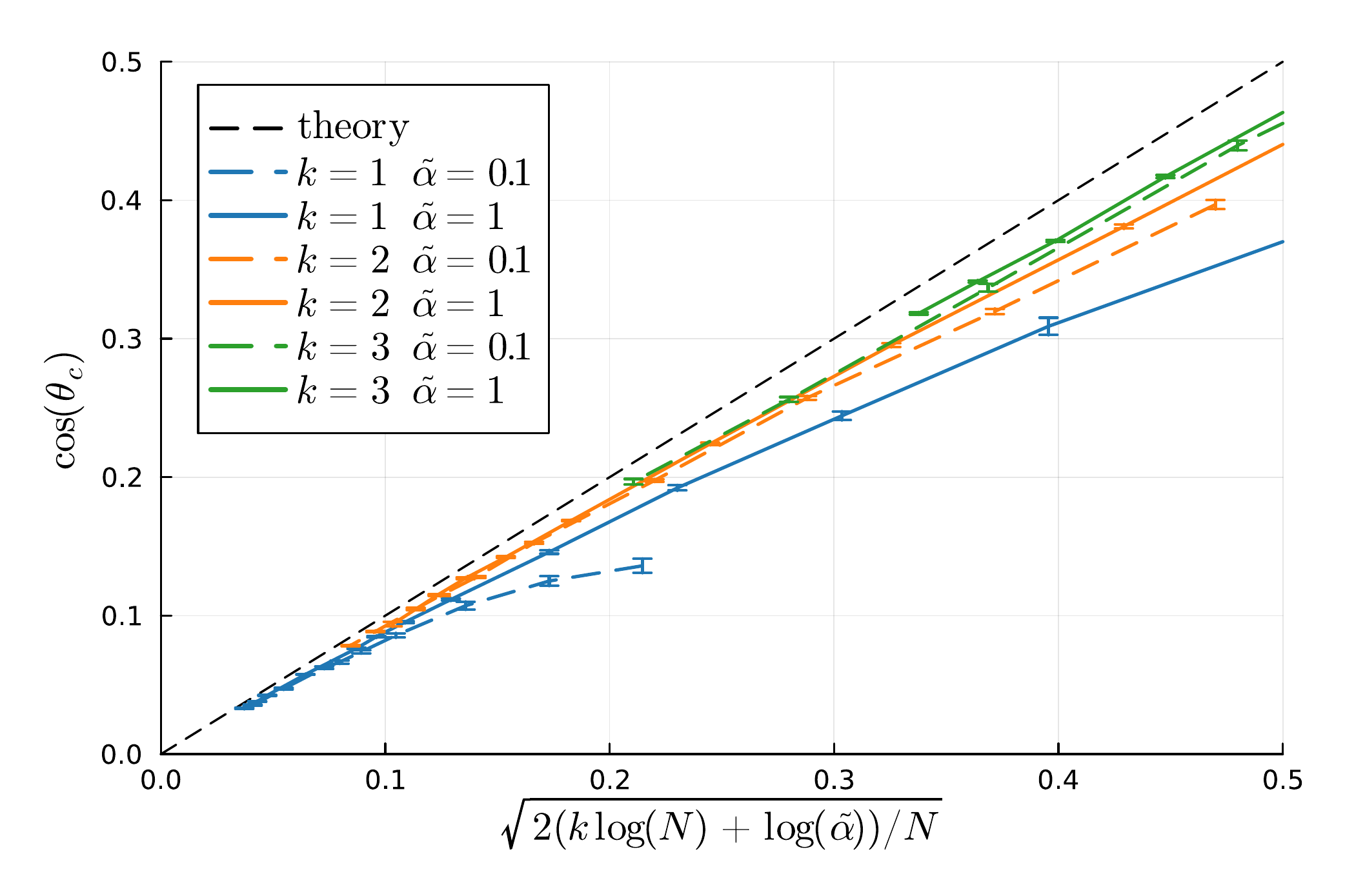}
	\caption{Size of attraction basins for patterns uniformly distributed on the hypersphere. We study here the case in which the number of stored patterns is polynomial, $P=\tilde{\alpha} N^k$, far below its exponential storage capacity. Given a pattern $\bxi^1$, we compute the cosine of the average angle to the nearest other pattern. We plot the expectation value of this cosine versus the theoretical prediction  $
\sqrt{2\frac{k\log N +\log \tilde{\alpha}}{N}}$. In the large $N$ limit, there is good agreement between the numerical result and the prediction.
\label{fig:basins_pol}}
\end{figure}

It is interesting to study the case of small $\alpha$, i.e. $\log P\ll N$.
 We can expand the rate function $s_\rho(\epsilon)$ close to its minimum.This can be obtained by expanding $\zeta-\rho(\lambda)$ close to $\lambda=0$. 
Let us write
\begin{align}
    \zeta_\rho(\lambda)=a\lambda+\frac{1}{2}b \lambda^2+O(\lambda^3).
\end{align}
In general, $a$ and $b$ depend on the distribution of patterns; however, for rotational invariant distributions normalized so that $\mathbb{E} |\bxi|^2=N$, we have simply $a=0,b=1$. Then
\begin{align}
    s_\rho(\epsilon)=\frac{1}{2}\epsilon^2 +O(\epsilon^3).
\end{align}
The condensation energy is given by 
$
   \epsilon_*=\sqrt{2 \alpha}.
$
For any finite $\lambda$ the REM is condensed and one has 
$
   \Phi=\epsilon_*=\sqrt{2 \alpha}=\sqrt{2 \frac{\log P}{N}}.
$
The angle $\theta_c$ characterizing the attraction basin is obtained from $\rho\cos\theta_c= \Phi$. This gives:
\begin{align}
    \cos\theta_c= 
\sqrt{\frac{2\log P}{N}}.
\end{align}
Let us assume that the above small $\alpha$ limit is valid to describe the case of polynomial storage, $P=\tilde{\alpha}\, N^k$. Then the prediction for the size of the attraction basins is 
\begin{align}
    \cos\theta_c= 
\sqrt{2\,\frac{k\log N +\log \tilde{\alpha}}{N}}.
\end{align}
We see that $\cos \theta_c$ goes to zero proportionally to $\sqrt{\log N/N}$. Therefore, using the dense associative memory network at a storage capacity that is much smaller than its (exponentially large ) threshold leads to a very large basin of attraction. 
We have tested this prediction in the case of spherical patterns. The results, presented in Fig. \ref{fig:basins_pol}, are in very good agreement with the theory.

\section{Spherical patterns}
\label{app:spherical}
\subsection{Typical case}
Consider the case where patterns $\bxi^\mu\in \mathbb{R}^N$ are uniformly distributed on the sphere $\lVert\bxi^\mu\rVert^2=N$, and choose a vector $\bx$ such that $\lVert\bx\rVert^2=\rho^2 N$. Using rotational invariance,  the distribution of the energy levels $\ve^\mu=\bx\cdot\bxi^\mu / N$ depends on $\bx$ only through $\rho$. One can choose the axes so that $\ve^\mu=\rho\xi^\mu_1/\sqrt{N} $. The probability density of these energy levels is expressed by the large deviation function
\begin{equation}
	s_\rho(\ve)=-\lim_{N\to\infty}\frac{1}{N}\log\frac{\int \prod_i d\xi_i\  \delta\left(N - \sum_i \xi_i^2\right)\,\delta\left(\ve -\frac{ \rho \xi_1}{\sqrt{N}}\right)}{\int \prod_i d\xi_i\  \delta\left(N - \sum_i \xi_i^2\right)}.
\end{equation}
As prescribed in Appendix \ref{app:rem3-summary},  we first compute its Legendre transform 
\begin{equation}
	\zeta_\rho(\lambda)=\lim_{N\to\infty}\frac{1}{N}\log\frac{ \int \prod_i d\xi_i\  \delta\left(N - \sum_i \xi_i^2\right) e^{\lambda \rho \xi_1\sqrt{N}}}{\int \prod_i d\xi_i\  \delta\left(N - \sum_i \xi_i^2\right)}.
\end{equation}
Standard field theoretical calculations and saddle point evaluation lead to 
\begin{equation}
	\zeta_\rho(\lambda)=  \frac{1}{2}\left(
	\sqrt{1+4\lambda^2 \rho^2}-1 -
	\log  \left(\frac{1+\sqrt{1+4\lambda^2 \rho^2}}{2} \right)
	\right)\ .
\end{equation}
Performing the inverse Legendre transform one gets
\begin{align}
	s_\rho(\ve) &=-
	\frac{1}{2}\log\left(1-\frac{\ve^2}{\rho^2}\right).
\label{eq:s-spherical}
\end{align}
The function $s_\rho(\ve)$ is shown in Fig. \ref{fig:ratefunction}.
According to Eq. \eqref{eq:phi_rem-final}, the REM free-energy density is given by
\begin{align}
	\phi_{\alpha,\rho}(\lambda) = 
\begin{cases} 
 \frac{\alpha}{\lambda} +\frac{1}{2\lambda}\left(
	\sqrt{1+4\lambda^2 \rho^2 }-1 -
	\log  \left(\frac{1+\sqrt{1+4\lambda^2 \rho^2 }}{2} \right])
	\right) &\lambda<\lambda_*(\alpha,\rho)\nonumber\\
\rho \sqrt{1-e^{-2\alpha}} & \lambda\geq\lambda_*(\alpha,\rho),
\end{cases} 
 \label{phirem_sph}
\end{align}
where 
\begin{align}
	\lambda_*(\alpha,\rho)= \frac{1}{\rho} e^{2\alpha}\sqrt{1-e^{-2\alpha}}.
\end{align}
According to Eq. \eqref{eq:alpha1}, the phase boundary for the retrieval of a typical pattern, $\alpha_1(\lambda)$, is found  by using the expression for $\phi_{\alpha,\rho}$ with $\rho=1$ (which is the radius of any pattern). It is given by the value of $\alpha$ satisfying that $\phi_{\alpha,1}(\lambda)=1$. This is plotted in Fig. \ref{fig:sph-pd}. The retrieval phase boundary is entirely in the region of the REM which is uncondensed.

\subsection{All patterns retrieval}

In order to study the retrieval of all patterns using the union bound, we notice that, for the spherical model, all patterns have $r=\lVert\bxi^1\rVert/\sqrt{N}=1$. Therefore, eq. \eqref{Adef} reduces to 
\begin{align}
A(\alpha,\lambda) = \inf_{\phi\,:\phi>1}I_{\alpha,1,\lambda}(\phi)
\end{align}
Assuming $\alpha < \alpha_1(\lambda)$, by the very definition of $\alpha_1$ we have that the typical REM free energy $\phi_{\alpha,1}(\lambda)$ is lower than 1. Therefore, we should look at atypically high free energy. According to Eq. \eqref{eq:I-largedev} we have
\begin{equation}
A(\alpha,\lambda) = \inf_{\phi\,:\phi>1} s_{1}(\phi)-\alpha.
\end{equation}
On the other hand, Eq. \eqref{eq:s-spherical} implies that $s_{1}(\phi)$ is finite only for $\phi < 1$. Therefore $A(\alpha,\lambda)=+\infty$ for $\alpha <\alpha_1(\lambda)$ and $A(\alpha,\lambda)=0$ otherwise. This leads to the final result $\alb(\lambda)=\alpha_c(\lambda)=\alpha_1(\lambda)$ for spherical patterns.

\subsection{Attraction basins}
The attraction basins are studied by considering an initial condition $\bx$ on the sphere $\lVert\bx\rVert^2=N$ sampled uniformly at random but at fixed angle $\cos(\theta)=\bx\cdot \bxi^1/ N$  from a typical pattern ($\bxi^1$ without loss of generality). As argued in Sec. \ref{sec:Basins}, the gradient flow (with or without projection on the sphere) will converge to $\bxi^1$ as long as the initial angle is smaller than some critical angle $\theta_c(\alpha,\lambda)$, that is  $\cos(\theta) > \cos(\theta_c)$. 
The critical angle is found solving $\phi_{\alpha,1}(\lambda)=\cos(\theta_c)$ as suggested by the energy decomposition \eqref{eq:ene_decomp}. The critical lines are shown in Fig. \ref{fig:basin_spherical}. For a given $\alpha$, the largest basin size is obtained for $\lambda \geq  e^{2\alpha}\sqrt{1-e^{-2\alpha}}$, where the critical angle becomes $\ctc(\alpha) = \sqrt{1 - e^{-2\alpha}}$. 

\section{Gaussian patterns}
\label{app:gaussian}
\subsection{Typical case}
Consider the case where patterns have independent identically distributed components which are Gaussian variables of mean zero and unity variance,
and independently choose a vector $\bx$ such that $\lVert\bx^2\rVert=N \rho^2$. Using rotational invariance,  the distribution of the energy levels $\ve^\mu=\bx\cdot\bxi^\mu / N$ depends on $\bx$ only through its norm $\rho$. One can choose the axes so that $\ve^\mu=\rho\xi^\mu_1/\sqrt{N} $. The probability density of this energy is expressed by the large deviation function
\begin{equation}
	s_\rho(\ve)=-\lim_{N\to\infty}\frac{1}{N}\log\left( \int \frac{d\xi_1}{\sqrt{2\pi}} \; e^{-\xi_1^2/2}\; \delta\left(\ve -\frac{ \rho \xi_1}{\sqrt{N}}\right)\right).
\end{equation}
The generating function is given instead by
\begin{equation}
		\zeta_\rho(\ve)=\lim_{N\to\infty}\frac{1}{N}\log\int \frac{d\xi_1}{\sqrt{2\pi}} \; e^{-\xi_1^2/2 + \frac{1}{\sqrt{N}}\lambda \rho \xi_1 },
\end{equation}
and the two functions are connected by a Legendre transform as described in Appendix 
\ref{app:rem1}.

Computing the two limits obtain
\begin{align}
	s_\rho(\ve)&=\frac{\ve^2}{2 \rho^2}\\
	\zeta_\rho(\lambda)&=\frac{\lambda^2 \rho^2}{2}.
\end{align}
The REM free-energy density, $\phi_{\alpha,\rho}(\lambda)$, according to Eq. \eqref{eq:phi_rem-final} reads
\begin{align}
\phi_{\alpha,\rho}(\lambda) = \begin{cases}
\frac{\alpha}{\lambda}+\frac{1}{2}\lambda \rho^2 & \lambda<\lambda_*(\alpha,\rho)\\
\rho\sqrt{2\alpha} & \lambda\geq\lambda_*(\alpha,\rho),
\end{cases}
\end{align}
where the condensation threshold is given by
\begin{align}
	\lambda_*(\alpha,\rho)= \frac{\sqrt{2\alpha}}{\rho}.
\end{align}

The phase boundary for retrieval of a typical pattern, $\alpha_1(\lambda)$, is found  by using the expression for $\phi_{\alpha,\rho}$ with $\rho=1$ (which is the radius of a typical pattern) and finding the value of $\alpha$ such that $\phi_{\alpha,1}(\lambda)=1$. 
It takes the simple form
\begin{equation}
	\alpha_{1}(\lambda) = \begin{cases}
		\lambda(1-\frac{1}{2}\lambda) &0\leq \lambda <1 \\
		\frac{1}{2} & \lambda\geq 1.
	\end{cases}
\end{equation}
The  phase diagram is plotted in Fig. \ref{Fig-gauss-pd}.

\subsection{All patterns retrieval}

We now compute the quantities involved in the all patterns retrieval lower bound $\alb(\lambda)$ given in Eq. \eqref{eq:alb}. We first compute the large deviation rate function $\tilde I(r)$ for the scaled norm of a single pattern,
$r=\lVert\bxi\rVert/\sqrt{N}$. It is easy to show that
\begin{align}
\tilde{I}(r)=-\lim_{N\to\infty} \frac{1}{N} \; \log\int \frac{d^N\bxi}{\sqrt{2\pi}^N} \;e^{-(1/2)\sum_i \xi_i^2} \; \delta\left(r-\frac{\lVert\bxi\rVert}{\sqrt{N}}\right)=\frac{1}{2}(r^2-1)-\log r.
\end{align}

Then we compute the large deviation function for the REM free energy conditioned on the pattern radius $r$. According to the general expression in Eq. \eqref{eq:I-largedev}, this is given by
\begin{equation}
	I_{\alpha,r,\lambda}(\phi)=\begin{cases}
		+\infty & \phi<\phi_{\alpha,r}(\lambda)\\
		0 & \phi=\phi_{\alpha,r}(\lambda)\\
		\frac{\phi^2}{2 r^2}-\alpha & \phi>\phi_{\alpha,r}(\lambda).
	\end{cases}
\end{equation}
For convenience , we define the function $I^*(\alpha,r,\lambda)$ such that Eq. \eqref{Adef} becomes:
\begin{align}
A(\alpha,\lambda) =\inf_{r\in[0,\infty)}\,\tilde{I}(r)+ I_*(\alpha,r,\lambda)
,
\end{align}
that is
\begin{equation}
   I_*(\alpha,r,\lambda)= \inf_{\phi\,:\phi>r^{2}}I_{\alpha,r,\lambda}(\phi).
   \label{mdef}
\end{equation}

Let us denote by $r_0(\alpha,\lambda)$ the value of $r$ such that $r^2=\phi_{\alpha,r}(\lambda)$. It is given by
\begin{equation}
r_0(\alpha,\lambda)=\begin{cases} 
\sqrt{\frac{\alpha}{\lambda(1-\lambda/2)}} & 0 \leq \lambda < 1\\
\sqrt{2\alpha} & \lambda \geq 1.
\end{cases}
\end{equation}
When $r<r_0$, we have $ \phi_{\alpha,r}(\lambda)>r^2$. When $r>r_0$ instead, we have $ \phi_{\alpha,r}(\lambda)<r^2$. It follows that:
\begin{itemize}
    \item If  $r>r_0$ the infimum in (\ref{mdef}) is obtained at $\phi=r^2$ and its value is $I_*(\alpha,r,\lambda)=\frac{r^2}{2}-\alpha$.
    \item If $r<r_0$  the infimum in (\ref{mdef}) is obtained at $\phi=\phi_{\alpha,r}(\lambda)$ and its value is $I_*(\alpha,r,\lambda)=0$.
\end{itemize}
This leads to 
\begin{equation}
A(\alpha,\lambda)= \min\left(
\min_{r\in[0,r_0(\alpha,\lambda)]}
\,\tilde I (r),\ 
\min_{r\in[r_0(\alpha,\lambda),\infty)}\,\tilde I (r)+\frac{r^2}{2}-\alpha\right).
\end{equation}
The final value for $\alb(\lambda)$ is found by solving $A(\alpha,\lambda)=\alpha$. 
When $\lambda \gtrsim 0.70091$ one gets $\alb=\log(2)/4$. We  have $\alb \sim \lambda$ instead for small $\lambda$, therefore in this limit it matches $\alpha_1(\lambda)$. The results are shown in Fig. \ref{Fig-gauss-pd}.

\subsection{Attraction basins and typical distances}
In the case of Gaussian patterns, the condition $\rho\, r_\xi\, \cos(\theta) > \phi_{\alpha,\rho}(\lambda)$ identifying the attraction basin of a typical pattern,  for large $\lambda$ takes the form $\cos(\theta_c) > \sqrt{2\alpha}$. Clearly, the basin disappears when $\alpha>0.5$

Since in this ensemble we have large deviations also in the patterns' radii, contrary to the spherical case, we can perform a slightly more refined version of the computation in Appendix \ref{app:typdist} in order to compute also the radius of the patterns with maximum overlap with a reference configuration.

Consider a reference configuration $\bx$ with squared norm $\lVert\bx\rVert^{2}=N$.
We ask what is the number of patterns with a certain overlap $c$
with it and with a certain square norm $r^2 N$. That is, we want to compute
the statistics of the random variable

\begin{equation}
	\mathsf{N}(c,r)=\sum_{\mu=2}^{P}\delta\left(\bx \cdot\bxi^{\mu}-cN\right)\,\delta\left(\lVert\bxi^{\mu}\rVert^{2}-r^2N\right)
\end{equation}

Although we are interested in the quenched average, $\frac{1}{N}\mathbb{E}\log\mathsf{N}(c,r)$, let's do an annealed computation first.

\begin{align}
	\mathbb{E}\mathsf{N}(c,r) & =e^{\alpha N}\mathbb{E}\int\frac{d\hat{c}}{2\pi}\,\frac{d\hat{r}}{2\pi}\ e^{+i\hat{c}cN+\frac{1}{2}\hat{r}r^2N-i\hat{q}\sum_{i}x_i\xi_{i}-\frac{1}{2}\hat{r}\sum_{i}\left(\xi_{i}\right)^{2}}\\
	& =\int\frac{d\hat{c}}{2\pi/N}\,\frac{d\hat{r}}{2\pi/N}\ e^{\alpha N+i\hat{c}cN+\frac{1}{2}\hat{r}r^2N-\frac{\hat{c}^{2}}{2(1+\hat{r})}N-\frac{1}{2}N\log(1+\hat{r})}\\
	& =\int\frac{d\hat{r}}{\sqrt{2\pi(1+\hat{r})/N}}\ e^{\alpha N+\frac{1}{2}\hat{r}r^2-\frac{1}{2}c^{2}(1+\hat{r})N-\frac{1}{2}N\log(1+\hat{r})}
\end{align}

Therefore

\begin{equation}
	\lim_{N\to+\infty}\frac{1}{N}\log\mathbb{E}\mathsf{N}(c,r)=\alpha+\frac{1}{2}\st_{\hat{r}}\ \left[\hat{r}r^2- c^{2}(1+\hat{r})-\log(1+\hat{r})\right].
\end{equation}
At the stationary point
\begin{equation}
	\hat{r}=\frac{1}{r^2-c^{2}}-1.
\end{equation}
This leads to 
\begin{equation}
	\mathrm{S}^{ann}_\alpha(c, r)=\lim_{N\to+\infty}\frac{1}{N}\log\mathbb{E}\mathsf{N}(c,r)=\alpha +\frac{1-r^2}{2}+\frac{1}{2}\log\left(r^2-c^2\right)
 \end{equation}

A standard second moment argument then gives the concentration of the quenched entropy:
\begin{equation}
S_\alpha(c,r)=\lim_{N\to\infty}\frac{1}{N}\log
\mathrm{N}(c)=
\begin{cases}
    \mathrm{S}^{ann}_\alpha(c,r) & \mathrm{S}^{ann}_\alpha(c,r)\geq 0\\
    -\infty &  \mathrm{S}^{ann}_\alpha(c, r)< 0
\end{cases}
\end{equation}

For a given overlap $c$, the radius of the patterns giving the dominant contribution to the energy is found by setting $\partial_r \mathrm{S}^{ann}_\alpha(c,r) = 0$. This gives 
\begin{align}
    r_*(c) &= \sqrt{1 + c^2},\\
    \mathrm{S}^{ann}_\alpha(c)&=\mathrm{S}^{ann}_\alpha(c,r_*(c)) = \alpha - \frac{1}{2}c^2 
\end{align}

Imposing the condition $\mathrm{S}^{ann}_\alpha(c,r_*(c)) = 0$ we finally recover the result $c_{max}(\alpha) = \sqrt{2\alpha}$.

\section{Storage capacity lower bound from Ramsauer et al. '20}
\label{app:ramsauer}
In this Appendix, we restate Theorem 3 of Ref. \cite{ramsauer2021hopfield} using our notation and we compute its asymptotic form in the high-dimensional limit $N\to+\infty$ with $P=e^{\alpha N}$ for fixed $\alpha$. The Theorem considers patterns $\bxi^\mu$ uniformly distributed on the sphere of radius $K\sqrt{N-1}$ and provides a lower bound for all patterns retrieval capacity at finite $N$ described in the following. We consider $K=1$ for simplicity and assume failure
probability $0<p\leq1$ for the storage problem, that is the probability that at least one pattern is not an approximate fixed point of the dynamics (we refer the reader to Ref. \cite{ramsauer2021hopfield} for precise definitions). We define the quantities 

\begin{align}
	a :=\frac{2}{N-1}\left(1+\log(2\lambda p(N-1))\right);\quad
	b  :=\frac{2}{5}\lambda;\quad
	c  :=\frac{b}{W_{0}(e^{a+\log b})};
\end{align}
where $W_{0}$ is the upper branch of the Lambert $W$ function. Ensuring
$c\geq\left(\frac{2}{\sqrt{p}}\right)^{\frac{N-1}{4}}$, with probability
$1-p$ the number of patterns that can be stored $P_c$ satisfies the bound \cite{ramsauer2021hopfield}
\begin{equation}
	P_c >\sqrt{p}c^{\frac{N-1}{4}}.
 \label{Ramsauer_inequality}
\end{equation}
We now consider the thermodynamic limit $N,P\to \infty$ and $p\to 0$ with:
\begin{align}
	p & =e^{-\gamma N},\\
	P_c & =e^{\alpha_c N}.
\end{align}
The limit for the previously defined quantitis is then
\begin{align}
	a =-2\gamma;\quad
	b =\frac{2}{5};\lambda\quad
	c =\frac{2}{5}\lambda\frac{1}{W_{0}\left(e^{-2\gamma+\log\left(\frac{2}{5}\lambda\right)}\right)}.
\end{align}
The inequality \eqref{Ramsauer_inequality} becomes
\begin{equation}
	\alpha_c >-\frac{1}{2}\gamma+\frac{1}{4}\log c.
\end{equation}
Taking the limit $\gamma\to 0$ we finally derived the asymptotic form of the bound presented in Ref. \cite{ramsauer2021hopfield}:
\begin{equation}
	\alpha_{c}(\lambda)>\frac{1}{4}\log \left[
 \frac{2\lambda/5}{W_{0}(2\lambda/5)} 
 \right]\eqqcolon \albr(\lambda)
\end{equation}
For large $\lambda$ we have $c \sim \frac{2\lambda}{5\log(\frac{2}{5}\lambda)}$ and $\albr \sim \frac{1}{4}\log\lambda$. For small $\lambda$ instead we have  $\albr \sim \lambda/10$.
The value of $\albr$ is plotted in Fig. \ref{fig:sph-pd}. One can notice a large gap with the exact value obtained by our large deviation analysis.

\section{Numerical Experiments}
\label{app:numerical}
This Appendix is devoted to the numerical validation of our analytical result for the typical pattern retrieval threshold $\alpha_1(\lambda)$ (more precisely, we consider its inverse, $\lambda_1(\alpha)$) and for the typical basin size. We make the general remark that while the numerical results are in good agreement with theory predictions for infinite $N$, dealing with an exponential number of patterns makes the system hard to simulate. We quickly saturate our memory and compute resources going up in $N$ so that is hard to extrapolate the large $N$ behavior. Therefore, the numerical validation of our theory cannot be made entirely satisfactory, and we hope that follow-up works will be able to establish with mathematical rigor our results obtained through partially non-rigorous arguments. 

In the numerical experiments, we perform a gradient descent (GD) procedure on the energy function of Eq. \eqref{eq:enedef}. We use a step size $\eta=0.5$. The results are averaged over a number of realizations of the patterns that vary from $1000$ (low values of $N$ and $\alpha$) to 10 (large $N$ and $\alpha$).

\subsection{Typical pattern retrieval}
\label{app:num-typical}
\begin{figure}[!htb]
	\includegraphics[width=\textwidth]{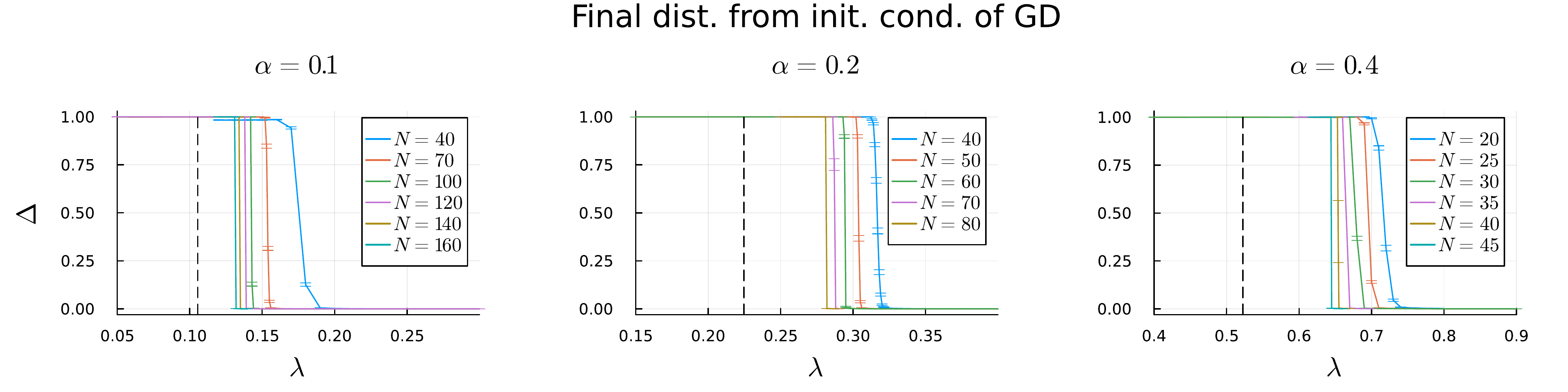}
	\caption{Spherical patterns. End distance $\Delta$ from initial conditions $\bxi^1$ under GD dynamics as a function of the interaction strength $\lambda$ and for different values of $N$ and $\alpha$. Vertical lines are the predicted values for the threshold $\lambda_1(\alpha)$ at infinite $N$ separating the retrieval and the non-retrieval regime.}
	\label{fig:gd_spherical}
\end{figure}

\begin{figure}[!htb]
	\includegraphics[width=\textwidth]{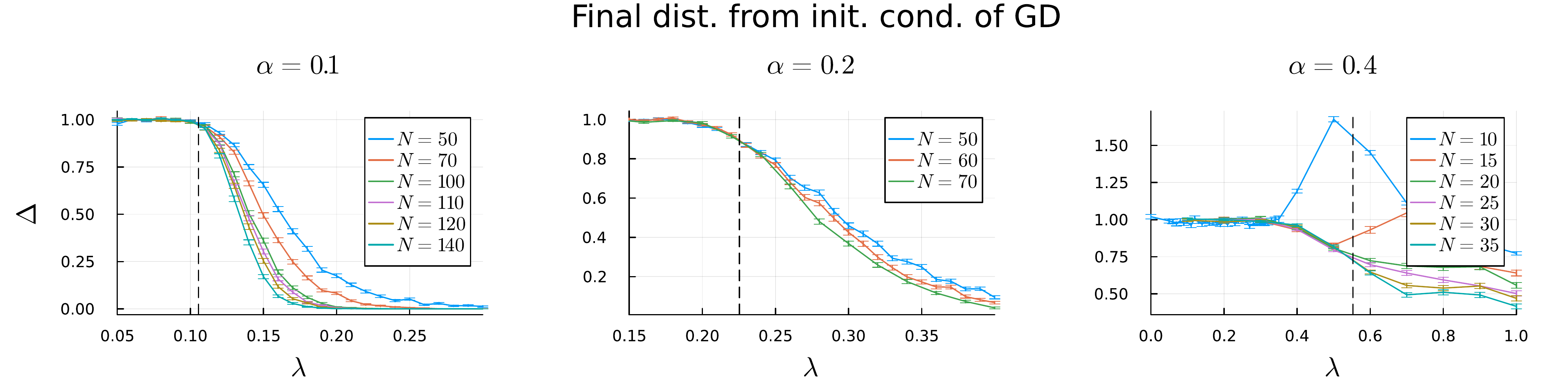}
	\caption{Gaussian patterns. End distance $\Delta$ from initial conditions $\bxi^1$ under GD dynamics as a function of the interaction strength $\lambda$ and for different values of $N$ and $\alpha$. Vertical lines are the predicted values for the threshold $\lambda_1(\alpha)$ at infinite $N$ separating the retrieval and the non-retrieval regime.}
	\label{fig:gaussian_GD}
\end{figure}

In the first set of experiments, we start from the initial condition $\bx^{t=0}=\bxi^1$ and run GD until convergence. We then plot the normalized distance from the initial configuration, $\Delta =\lVert\bx^{t=\infty} - \bxi^1\rVert^2 / N$, as a function of $\lambda$ and for different values of $N$ and $\alpha$. The results are presented in Fig. \ref{fig:gd_spherical} for the spherical case and in Fig. \ref{fig:gaussian_GD} for the Gaussian case. In both cases, at large $\lambda$ GD remains close to the initial point, meaning that the pattern is an (approximate) minimum. For low $\lambda$ GD approaches the origin instead, a phenomenon already outlined in Ref. \cite{ramsauer2021hopfield} as the "average of all patterns" global fixed point.
As expected, the crossover between the two regimes becomes sharper as $N$ increases and approaches the analytical prediction for the threshold $\lambda_1(\alpha)$ obtained by inverting Eq. \eqref{eq:alpha1}. We notice that the transition is much smoother in the Gaussian case compared to the spherical case. 

In Fig. \ref{fig:sph-pd}  (bottom), we present a large $N$ extrapolation in the case of spherical patterns. At given $\alpha$ and $N$, we define the crossover value of $\lambda$ as the smallest value of $\lambda$ for which $\Delta < 0.5$. We then extrapolate the numerical results to large $N$ through a quadratic fit in $\frac{1}{N}$ and compare them with the analytical prediction $\lambda_1(\alpha)$.

\subsection{All patterns retrieval for large $\lambda$}

In the case of Gaussian patterns, we have a gap between the all patterns retrieval lower bound $\alb(\lambda)$ of Eq. \eqref{eq:alb}  and the single pattern one $\alpha_{1}(\lambda)$, that is $\alb(\lambda) < \alpha_{1}(\lambda)$. This can be clearly seen in the phase diagram shown in Fig. \ref{Fig-gauss-pd}.
It remains to be established  where the true value of the all patterns retrieval threshold $\alpha_c(\lambda)$ lies in the interval $[\alb(\lambda), \alpha_{1}(\lambda)]$ and if the gap really exists. Since this question cannot be answered within the theory we have developed, we provide some indications through numerical simulations.

Due to computational constraints at large $N$, we only explore the case of large $\lambda$ here and drop the explicit $\lambda$ dependence from the thresholds $\alpha_{1}, \alb, \alpha_c$.
For a given value of  $\alpha$ and for a given realization of the patterns, we check if all patterns in the sample are stable (i.e. correspond to a local minimum of the energy). As $\alpha$ grows, the probability $p^{\text{allretr}}_N(\alpha)$ of this event will drop to zero in correspondence of $\alpha_c$. 
The probability of the event "all patterns are retrieved", in the large $\lambda$ case is defined by
\begin{equation}
	p^{\text{allretr}}_N(\alpha) = \mathbb{E}\  \mathbb{I}\left(\lVert\bxi^\mu\rVert^2 > \bxi^\mu \cdot \bxi^\nu \  \forall\mu,\nu\in[P]\right).
\end{equation}
This is to be compared with the probability of retrieving a typical pattern, that is the probability that most patterns are retrieved. This can be estimated by computing the expected fraction of retrieved patterns
\begin{equation}
	f^{\text{retr}}_N(\alpha) =\mathbb{E} \frac{1}{P} \sum_{\mu=1}^P \mathbb{I}\left(\lVert\bxi^\mu\rVert^2 > \bxi^\mu \cdot \bxi^\nu \  \forall\nu\in[P], \nu\neq\mu\right).
\end{equation}
For large $N$, the fraction $f^{\text{retr}}_N(\alpha)$ should go from 1 to 0 at $\alpha_1$. 

\begin{figure}[!htb]
	\includegraphics[width=0.9\textwidth]{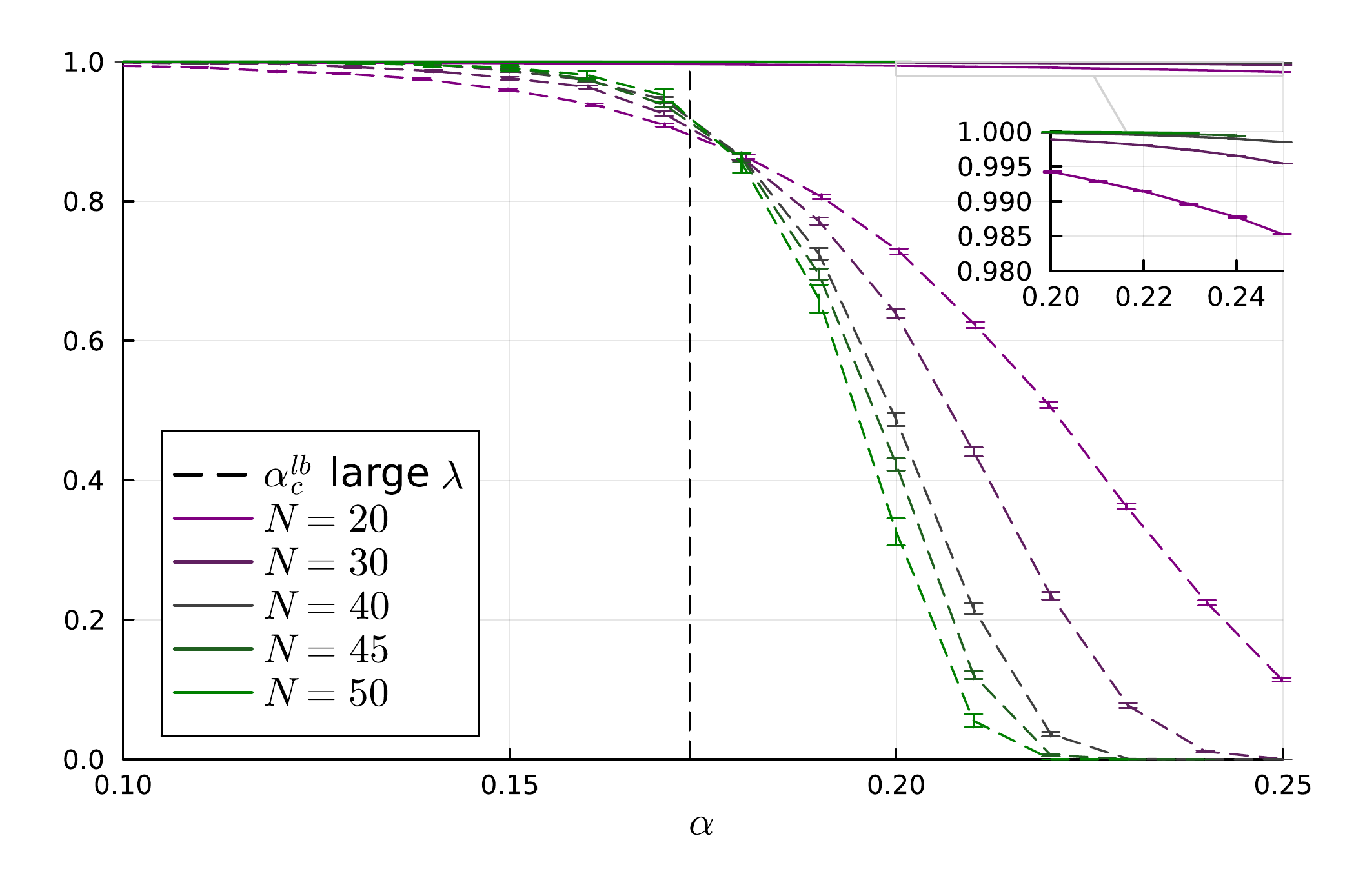}
	\caption{Gaussian patterns ensemble. (Dashed Lines) Numerical estimations of the probability of retrieving all patterns $P=e^{\alpha N}$ as a function of the load $\alpha$, for large $\lambda$ and for different system sizes $N$. Vertical dashed line is the lower bound of Eq. \eqref{eq:alb} . (Continuous Lines) Fraction of patterns retrieved, also shown in the inset, converging to 1 at large $N$ in the range explored as expected since $\alpha <\alpha_1$ .}
	\label{fig:pairwise_comparison}
\end{figure}

We estimate $p^{\text{allretr}}_N(\alpha)$ and $f^{\text{retr}}_N(\alpha)$ for different values of $N$ and $\alpha$ using Monte Carlo samples. The results are shown in Fig. \ref{fig:pairwise_comparison}.
The estimated value of $\alpha_c$, where we observe the crossover from 1 to 0 for $p^{\text{allretr}}_N(\alpha)$, seems to be close to the theoretically predicted $\alb=\log(2) / 4$ and far from $\alpha_1=0.5$. It remains an open question whether we have $\alb=\alpha_c$ or not.

\subsection{Basins}
In order to confirm our prediction $\cos \theta_c = \phi_{\alpha,1}(\lambda)$ for the basin size in the case of spherical patterns, we run the following experiments. 

\begin{figure}[!htb]
	\includegraphics[width=0.45\textwidth]{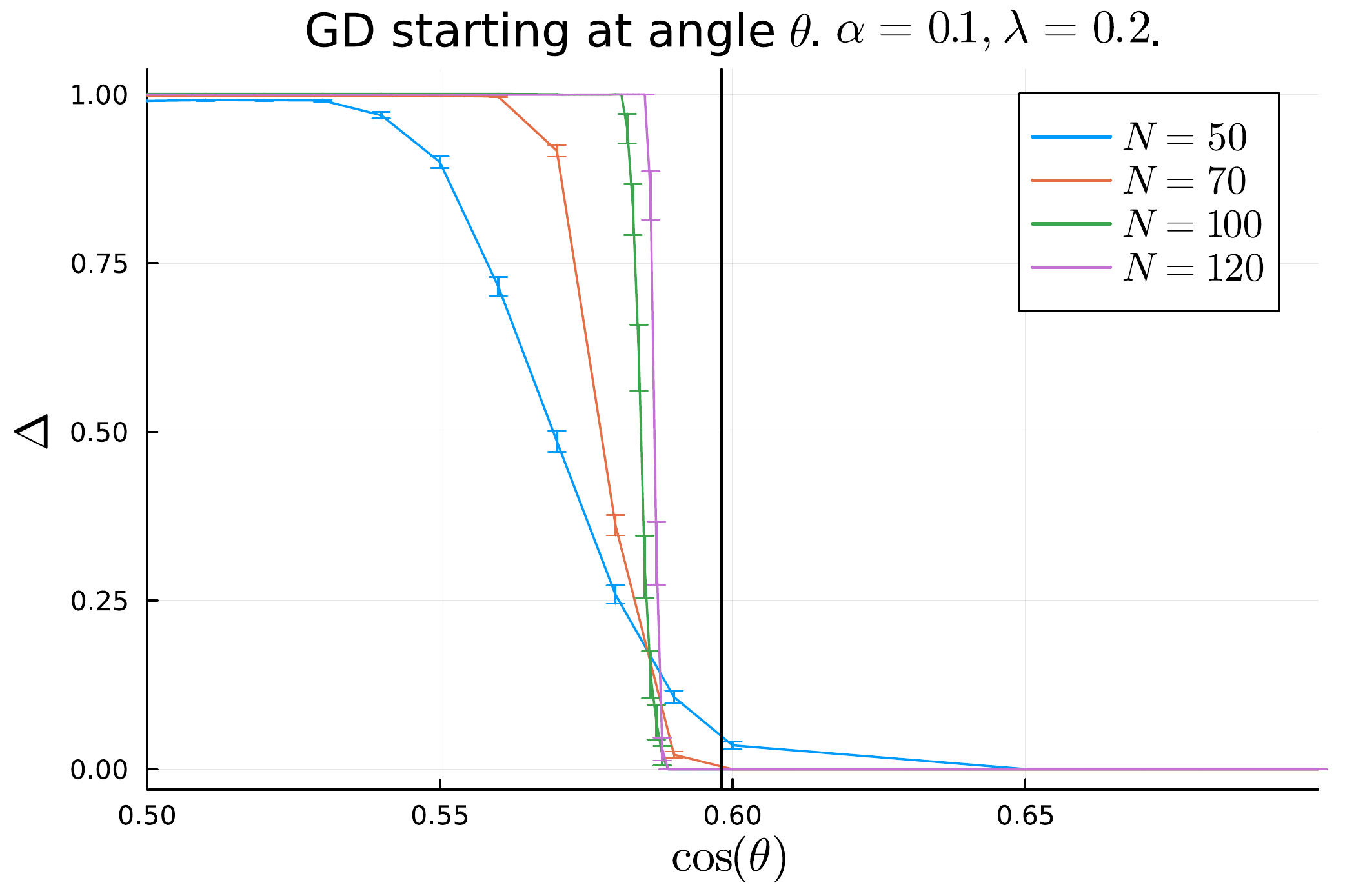}    
	\includegraphics[width=0.45\textwidth]{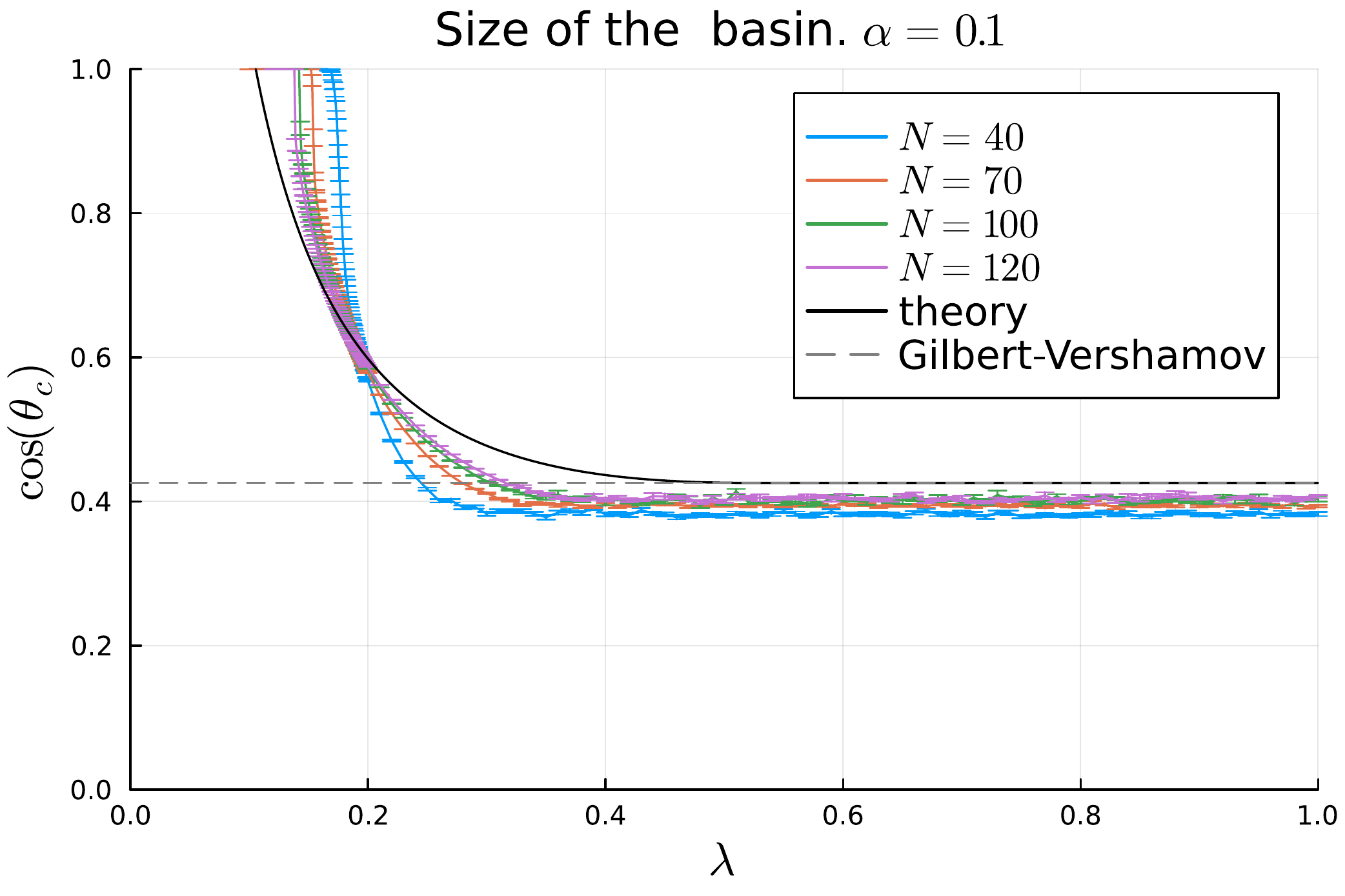}
	\caption{
 Spherical patterns.
 (Left) Distance $\Delta$ from the reference pattern $\bxi^1$ as a function of the starting angle at the end of GD dynamics. We set $\alpha=0.1$ and $\lambda=0.2$, with spherical patterns, and simulate different system sizes $N$. The theory prediction for the critical angle is given by the vertical line. (Right) Critical angle for the basin as a function of interaction strength. Numerical simulations at finite size are compared to the theory prediction for infinite size and to the Gilbert-Varshamov result for the angle of the nearest pattern.}
	\label{fig:num_basins}
\end{figure}

We sample an initial configuration for the GD dynamics uniformly at random on the hypersphere of radius $\sqrt{N}$ conditional on it being at a given angle $\theta$ from $\bxi$. We then measure the normalized distance of the final configuration from $\bxi^1$  after convergence of GD.
The result is shown in Fig. \ref{fig:num_basins} (Left) for $\alpha=0.1,\lambda=0.2$ and different values of $N$ as function of $\cos(\theta)$. For small angle ($\cos (\theta)$ close to 1), the final distance is close to 0, therefore the initial configuration lies in the basin of attraction of $\bxi^1$. Increasing the angle we have a crossover to a regime where the configuration escapes from $\bxi^1$. As $N$ increases the crossover becomes sharper and approaches the analytical prediction.

In Fig. \ref{fig:num_basins} (Right)  we plot the critical angle at finite size $N$ as a function of interaction strength. Since for finite $N$ we don't have sharp thresholds, we define the critical angle as follows: for a given sample, the critical angle is defined as the one for which half of the random initializations of the dynamics (we use 10 restarts) fall back to $\bxi^1$ while the other half escape the basin. We apply a bisection method to find such angle and then average the (cosine of the) angle over 10 samples. As the plot shows, increasing the system size the curves converge to the infinite size theoretical prediction. 

\subsection{Scaled dot-product}
\label{app:scaled}
\begin{figure}[!htb]
	\centering
	\includegraphics[width=\textwidth]{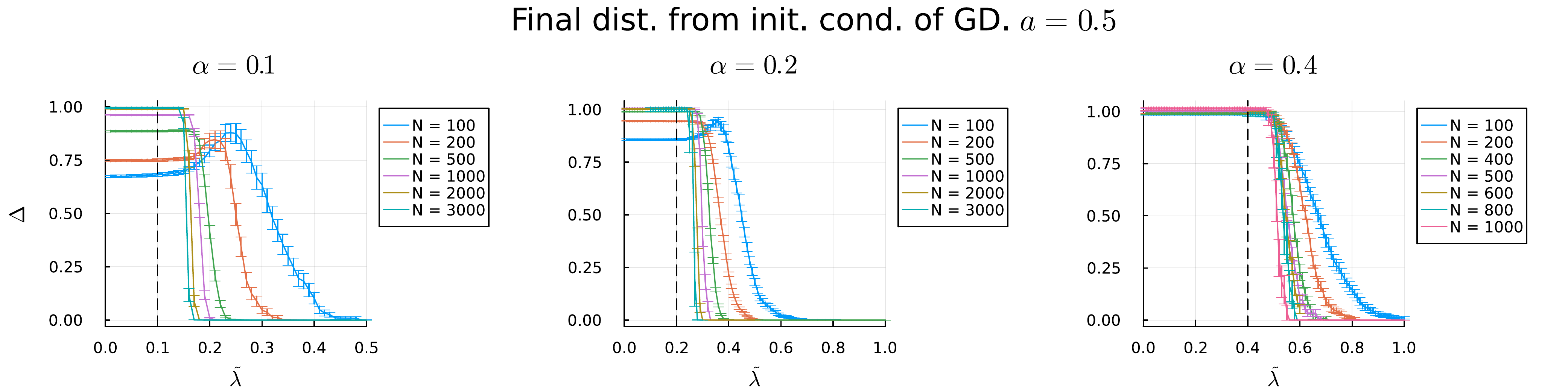}
	\includegraphics[width=\textwidth]{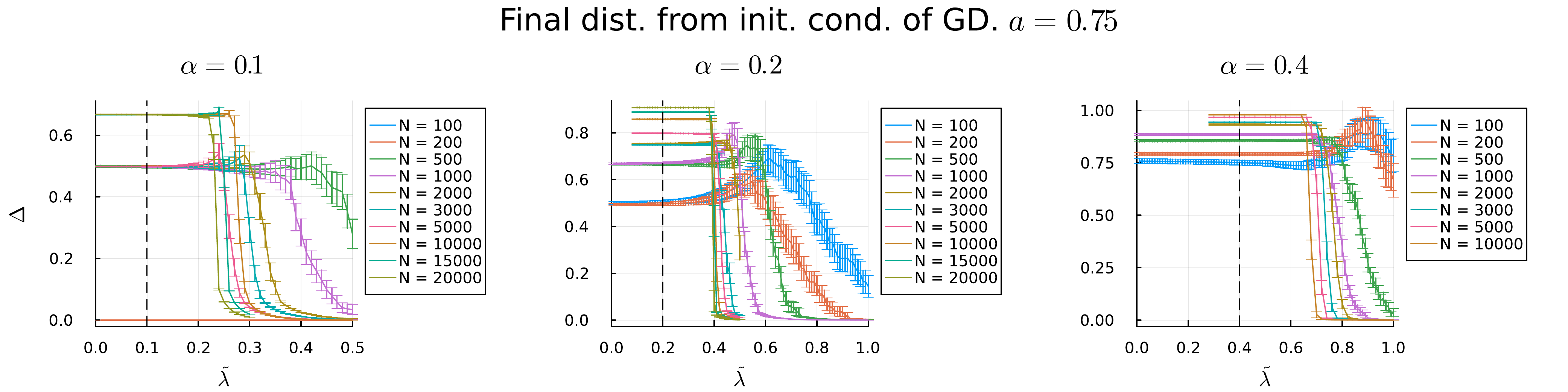}
	\caption{
 Gaussian patterns.
 Final distance from pattern $\bxi^1$ after GD for scaled attention with exponents $a=1/2$ (top) and  $a=3/4$ (bottom). Vertical lines are the theoretical predictions $\lambda_1(\tilde{\alpha}) = \tilde{\alpha}$ for the retrieval transition. The patterns are sampled from the Gaussian ensemble.}
	\label{fig:gaussian_gd_scaled}
\end{figure}

In Fig. \ref{fig:gaussian_gd_scaled} we present numerical validation for single pattern retrieval threshold $\tilde{\lambda}_1(\tilde{\alpha})$ for the scaling regime discussed in Section \ref{sec:scaled}
in the case of  Gaussian patterns. Here the number of patterns is $P = \exp(\tilde{\alpha} N^{1-a})$
and the energy reads
\begin{equation}
	E(\bx) = -\frac{N^a}{\tilde\lambda}\log \sum_\mu e^{\frac{\tilde\lambda}{N^a} \bx \cdot \bxi^\mu} + \frac{1}{2}\lVert \bx\rVert^2.
\end{equation}
The numerical protocol is the Gradient Descent one discussed in $\ref{app:num-typical}$. We observe that by increasing $N$ the retrieval crossover approaches the theoretical prediction as expected.

\end{document}